\documentclass{aa}
\usepackage[varg]{txfonts}

\usepackage{natbib}
\bibpunct{(}{)}{;}{a}{}{,}
\usepackage[colorlinks=true,citecolor=blue]{hyperref}

\usepackage{threeparttablex}
\usepackage{longtable}
\usepackage{lscape}

\usepackage{verbatim}
\usepackage{makecell}
\usepackage{caption}
\usepackage{subcaption}
\usepackage{multirow}
\usepackage{booktabs}
\usepackage{hyperref}

\def\kms{km\,s$^{-1}$}
\def\teff{$T_{\rm eff}$}
\def\logg{$\log\,g$}
\def\vmicro{$\xi$}

\begin{document}

\title{Observational mapping of the mass discrepancy in eclipsing binaries: Selection of the sample and its photometric and spectroscopic properties.\thanks{Based on observations collected at the European Southern Observatory, La Silla, Chile under program 0106.A-0906(A), and the HERMES spectrograph mounted on the KU Leuven Mercator telescope, La Palma, Spain.}}
\author{Andrew Tkachenko\inst{\ref{KUL}}
\and Kre\v{s}imir Pavlovski\inst{\ref{Zagreb},\ref{KUL}}
\and Nadezhda Serebriakova\inst{\ref{KUL}}
\and Dominic M. Bowman\inst{\ref{NU},\ref{KUL}}
\and Luc IJspeert\inst{\ref{KUL}}
\and Sarah Gebruers\inst{\ref{KUL},\ref{MPIA}}
\and John Southworth\inst{\ref{Keele}}
}

\institute{Institute of Astronomy, KU Leuven, Celestijnenlaan 200D, B-3001 Leuven, Belgium \\ \email{andrew.tkachenko@kuleuven.be} \label{KUL} 
\and Department of Physics, Faculty of Science, University of Zagreb, 10\,000 Zagreb, Croatia \label{Zagreb}
\and School of Mathematics, Statistics and Physics, Newcastle University, Newcastle upon Tyne, NE1 7RU, UK \label{NU}
\and Max Planck Institute for Astronomy, K\"onigstuhl 17, 69117 Heidelberg, Germany \label{MPIA}
\and Astrophysics Group, Keele University, Staffordshire ST5 5BG, UK \label{Keele}
}

\date{Received XXX / Accepted XXX}

\abstract{Eclipsing spectroscopic double-lined binaries are the prime source of precise and accurate measurements of masses and radii of stars. These measurements provide a stringent test of models of stellar evolution that are persistently reported to contain major shortcomings.} 
{The mass discrepancy observed for the eclipsing spectroscopic double-lined binaries is one of the manifestations of shortcomings in stellar evolution models. The problem reflects the inability of the models to accurately predict effective temperature and surface gravity or luminosity of the star for a given mass. Our ultimate goal is to provide an observational mapping of the mass discrepancy and propose a recipe for its solution.} 
{We initiate a spectroscopic monitoring campaign of 573 candidate eclipsing binaries classified as such based on their TESS light curves. In this work, we present a sub-sample of 83 systems for which orbital-phase resolved spectroscopy has been obtained and subsequently analysed with the methods of least-squares deconvolution and spectral disentangling. In addition, we employ TESS space-based light curves to provide photometric classification of the systems according to the type of their intrinsic variability.}
{We confirm 69 systems as either spectroscopic binaries or higher-order multiple systems. Twelve stars are classified as single and two more objects are found at the interface of their line profile variability being interpreted as due to binarity and intrinsic variability of the star. Moreover, 20 eclipsing binaries are found to contain at least one component that exhibits stellar oscillations. Spectroscopic orbital elements are obtained with the method of spectral disentangling and reported for all systems classified as either SB1 or SB2. The sample presented in this work contains both detached and semi-detached systems and covers a range in the effective temperature and mass of the star of \teff\ $\in [7000,30000]$~K and $M\in [1.5,15]$~M$_{\odot}$, respectively.}
{From the comparison of our own results with those published in the literature for well-studied systems, we conclude an appreciable capability of the spectral disentangling method to deliver precise and accurate spectroscopic orbital elements from as few as 6-8 orbital phase-resolved spectroscopic observations. Orbital solutions obtained this way are accurate enough to deliver age estimates with accuracy of 10\% or better for intermediate-mass F-type stars, an important resource for calibration of stellar evolution models for future space-based missions like PLATO. Finally, despite its rather small size relative to 573 system that we will ultimately monitor spectroscopically, the sample presented in this work is already suitable to kick off observational mapping of the mass discrepancy in eclipsing binaries.}


\titlerunning{Observational mapping of the mass discrepancy: selection and properties of the sample.}
\authorrunning{Tkachenko et al.}
\maketitle

\section{Introduction}\label{sec:introduction}
Eclipsing spectroscopic double-lined (hereafter eSB2) binaries are the prime source of precise and accurate mass measurements of stars, the parameter that, along with the initial composition of the star determines its evolutionary path and fate. Owing to the fact that their mass determination is largely model-independent, eSB2s represent an ideal test bed for the theory of stellar structure and evolution (SSE).

One of the striking outcomes of the SSE model test is the mass discrepancy. The term was introduced originally by \citet{Herrero2000} in the context of single stars, where the authors report a discrepancy between stellar masses inferred from the spectroscopic properties of stars and those found from the comparison of the position of stars in the Hertzsprung--Russell diagram (HRD) with evolutionary tracks. Although the authors conclude that the mass discrepancy is an observational manifestation of shortcomings in SSE models, it is worth noting that the spectroscopic masses are also model-dependent as they are inferred from the spectroscopic estimates of \logg\ and radii of the stars, both largely dependent on physics employed in the models of stellar atmospheres and winds. 

The conclusions regarding possible imperfections in SSE models are more sound when based on a comparison between model-independent masses inferred from the binary dynamics (dynamical masses hereafter) and those dictated by SSE models (evolutionary masses hereafter). Although earlier works suggest a good agreement between the dynamical and evolutionary masses for stars up to 25~M$_{\odot}$ \citep[e.g.][]{Burkholder1997}, more recent studies report a systematic discrepancy between the two types of measurement. Notably, two independent studies of the V380~Cyg system \citep{Guinan2000,Pavlovski2009b} found a large mass discrepancy for the more evolved primary component. While \citet{Guinan2000} claim the discrepancy can be resolved with an increased efficiency of the near-core mixing in the form of core overshooting, \citet{Pavlovski2009b} focus on the effect of rotation in SSE models and conclude that it cannot explain the observed mass discrepancy. Moreover, \citet{Tkachenko2014a} re-investigate the V380~Cyg system based on space-based photometric and high-resolution spectroscopic data and find the mass discrepancy for both binary components. The authors conclude that the discrepancy can only be resolved when the combined effect of rotation and core overshooting is included in SSE models.

The need for a more efficient near-core mixing becomes one of the streamlined theoretical hypotheses to explain the origin of the mass discrepancy in eclipsing binaries. \citet{Massey2012} and \citet{Morrell2014} argue the need for an increased efficiency of interior (rotational or core overshoot) mixing in SSE models to accommodate a small (some 10\%) but systematic discrepancy between dynamical and evolutionary masses in five high-mass eclipsing binaries in the LMC. The same conclusion is presented by \citet{Pavlovski2018} based on the study of four high-mass eclipsing binaries in the Galaxy. Furthermore, \citet{Rosu2020,Rosu2022a,Rosu2022b} invoke a simultaneous measurement of  the stellar parameters, rate of the apsidal motion and $k_2$ internal structure constant in the HD~152248, HD~152219, CPD-41{\textdegree} 7742, and HD~152218 binary systems to demonstrate the lack of efficiency of interior mixing in standard SSE models, in line with the conclusions drawn by \citet{Guinan2000} for the V380~Cyg system.

\citet{Claret_Torres2016,Claret_Torres2017,Claret_Torres2018,Claret_Torres2019} present a series of papers focusing on the inference of the efficiency of near-core mixing in the form of the convective core overshooting in a sample of some 50 eSB2 systems from \citet{Torres2010} and the catalogue of the Optical Gravitational Microlensing Experiment (OGLE)\footnote{\url{http://ogle.astrouw.edu.pl/}}. The authors consider two implementations of the core overshooting in Granada \citep{Claret2004,Claret2012} and {\sc mesa} \citep{Paxton2011,Paxton2013,Paxton2015,Paxton2018,Paxton2019} models, i.e. the convective penetration \citep{Zahn1991} and exponential diffusive approximation \citep{Freytag1996,Herwig2000}. Irrespective of the prescription used, the authors report a strong dependency of the inferred overshooting parameter on the stellar mass such that there is an almost linear transition from no overshooting to approximately $\alpha_{\rm ov}$ ($f_{\rm ov}$) = 0.2 (0.02)~$H_{\rm p}$ in the mass range from some 1.2 to 2.0~M$_{\odot}$. The authors find that the distribution flattens for stars more massive than 2.0~M$_{\odot}$. \citet{Viani2020} investigate an asteroseismic sample of stars observed with the {\it Kepler} mission in the mass range between 1.1 and 1.5~M$_{\odot}$. The authors report a strong positive correlation between the inferred overshooting parameter with the stellar mass. 

\citet{Costa2019} revisit the sample of \citet{Claret_Torres2016,Claret_Torres2017,Claret_Torres2018,Claret_Torres2019} using a combination of the Bayesian statistical framework and {\sc parsec} stellar evolution models. The authors confirm the need for a mild convective core overshooting for stars with masses $M\geq1.9$~M$_{\odot}$ and, additionally, find a large spread in the inferred $\alpha_{\rm ov}$ values ranging from some 0.3 to 0.8~$H_{\rm p}$. \citet{Costa2019} arrive at the conclusion that while the low boundary of the inferred $\alpha_{\rm ov}$ values is likely to be explained by an insufficient amount of core-boundary mixing in standard SSE models, the large spread in the inferred values is likely a manifestation of the natural distribution of initial rotation rates, and hence active rotational mixing in these intermediate-mass stars. 

Finally, \citet{Tkachenko2020} present a study of eleven high-mass eSB2 systems for which high-precision stellar quantities were derived in a homogeneous way \citep{Pavlovski2018, Pavlovski2023, Tkachenko2014a, Johnston2019b}. No dependence of the observed mass discrepancy on the stellar mass is reported but a strong anti-correlation with the surface gravity of the star is found. The same correlation persists when the stellar mass is replaced with the mass of the convective core of the star, the finding that the authors interpret as the need for larger convective core masses in standard models of SSE. Remarkably, \citet{Johnston2021} arrived at the same conclusion from the study of a combined large sample of binary components and asteroseismically active single stars. Furthermore, \citet{Martinet2021} investigate a large sample of stars from literature spanning a mass range from 7 to 25~M$_{\odot}$ and using a large grid of SSE models computed for different values of the overshooting parameter and initial rotation, and under the assumptions of a moderate and strong angular momentum transport in the models. The authors confirm the findings by \citet{Tkachenko2020} and \citet{Johnston2021} and report the need for larger convective cores at higher stellar masses.

An alternative hypothetical explanation of the mass discrepancy comes from the nature of binarity itself and the fact that the majority of high-mass stars are expected to be found in binary systems and/or have experienced binary interactions in the past \citep[e.g.,][]{Sana2012,Sana2014}. \citet{Mahy2020} investigate 26 eclipsing binary systems in the LMC and report a good agreement between their dynamical and spectroscopic masses whereas the evolutionary masses are found to be systematically overestimated. Upon a closer inspection of the obtained results, the authors find the mass discrepancy to be more pronounced in semi-detached binary configurations, the finding that suggests that binary interactions might be (at least partially) responsible for the observed mass discrepancy.

Despite being known for some three decades and many attempts made to observationally quantify and pinpoint the most likely theoretical explanation for it, the mass discrepancy persists as a phenomenon and the only definite statement the stellar astrophysics community can make about it is that the problem exists. The majority of previous studies focus on a particular mass regime and there is a large diversity in the employed data analysis methods and input physics used in SSE models, which makes it non-trivial to account for systematic uncertainties in an attempt to quantify and ultimately resolve the mass discrepancy problem. Motivated by the current state-of-the-art, we have initiated a systematic search for intermediate- to high-mass (spectral types OBAF and masses larger than some 1.2~M$_{\odot}$) eclipsing binaries in TESS \citep{Ricker2015} space-based photometric data and have organised a complementary ground-based spectroscopic follow-up campaign of the detected candidate eclipsing binaries. Our goal is to build a stellar sample that covers the entire mass range of stars born with the convective core, all the way up to some 30~M$_{\odot}$, and includes binaries with and without pulsating component and in different orbital configurations.

Our newly compiled sample is a generalisation of the binary sample presented in \citet{Tkachenko2020} to cover a much larger range of the stellar mass, account for intrinsic variability of stars in binaries, and for different orbital configurations such as detached versus semi-detached systems. We ultimately aim at a self-contained and homogeneous detailed analysis of all suitable eSB2, such that the analysis methods remain largely the same, thus minimising systematic uncertainties on the inferred parameters and observational mapping of the mass discrepancy across the HRD. In this paper, we present the sample selection of candidate eclipsing binary systems and provide an overview of the ground-based spectroscopic observations we have obtained so far (Section~\ref{sec:sample_selection}). We provide a more detailed look into the spectroscopic properties of the observed systems with the methods of least-squares deconvolution and spectral disentangling in Section~\ref{sec: spectroscopic_analysis}. Analysis of the TESS photometric data with the aim to provide a photometric classification according to the type of intrinsic variability of the sample stars is presented in Section~\ref{sec: photometric_analysis}. We proceed with a discussion of the obtained results in the context of the previous studies (Section~\ref{sec:discussion}) and close the paper with some concluding remarks and a list of future prospects (Section~\ref{sec:conclusions}).

\section{Sample selection}\label{sec:sample_selection}

Our sample selection is based on the work of \citet{IJspeert2021} that presents a magnitude limited catalogue of eclipsing binaries observed by the TESS mission. While a detailed description of all the analysis steps is provided in \citet{IJspeert2021}, here we provide a brief summary of relevance for the rest of the analysis in this work. 

\citet{IJspeert2021} start with a global all-sky selection of OB(A)-type stars from the TESS Input Catalogue \citep[version TICv8][]{Stassun2019}. The authors limit the selection to TESS magnitudes below 15 and exclude all targets that are resolved in the {\it Gaia} DR2 catalogue \citep{GaiaCollabBrown2018} but are recognised as a single object in the 2MASS catalogue. The 2MASS (J-K)--(J-H) colour-colour plane is employed to select some 205\,000 OB(A)-type candidate stars with the exact colour cuts informed by the OB-type star variability samples of \citet{Bowman2019} and \citet{Pedersen2019} \citep[cf. Figure~1 in][]{IJspeert2021}. In the next step, the authors exclude white dwarfs and giants from the sample using dedicated {\it Gaia} flags and limit themselves to stars for which either TESS-SPOC\footnote{DOI: \url{10.17909/t9-wpz1-8s54}} or MIT QLP\footnote{DOI: \url{10.17909/t9-r086-e880}} light curves could be retrieved from the MAST database\footnote{\url{http://archive.stsci.edu/tess/all_products.html}}. This step reduces the stellar sample of unique targets to some 91\,000. Ultimately, in a search for eclipsing binaries in the selected sample of OB(A)-type stars with TESS light curves, \citet{IJspeert2021} develop an algorithm for automated detection of eclipses and, if detected, determination of the binary orbital periods. The authors report 5502 candidate eclipsing binaries of which 2077 and 3425 are found to be false positives and true eclipsing binaries, respectively.

\subsection{Spectroscopic observations}

From the above-mentioned sample of 3425 eclipsing binaries, we select all targets with V magnitude below 11 and accessible from the Roque de los Muchachos Observatory (La Palma, Spain) in the course of the year. A total of 545 systems are proposed for observations with the High-Efficiency and high-Resolution Mercator Echelle Spectrograph \citep[{\sc hermes};][]{Raskin2011} instrument attached to the 1.2-m Mercator telescope, as part of a dedicated large programme. For each of those systems, we request eight epochs of spectroscopic observations with uniform sampling of the orbital phase. The latter is computed from the orbital period value derived by \citet{IJspeert2021} and a reference date $T_0$ taken to be the date of the first spectroscopic observation. A few systems have been followed up more extensively to resolve the effect of intrinsic stellar oscillations on spectral line profiles, with the time-series ranging from a couple of dozen spectra to almost 200 measurements (e.g. the case of 16~Lac in Table~\ref{Tab:targets}; see \citealt{Southworth2022a}).

At the time of writing, we have completed spectroscopic observations with the {\sc hermes} instrument for 58 systems from our sample. The criteria of completion are: (i) individual epoch observations meet the minimum signal-to-noise-ratio (S/N) requirement of 60 per resolution element, and (ii) a quasi-uniform orbital phase coverage to maximise the chances for detecting lines of a companion star in the observed composite spectra. The former requirement guarantees S/N in excess of 100 in the disentangled spectra of SB2 systems or the combined spectrum for SB1 binaries. Table~\ref{Tab:targets} gives an overview of the sample of 58 systems with columns 1 through 7 indicating a star name and a catalogue identifier (col. 1-3), orbital period (col. 4), number of spectra acquired (col. 5), and spectroscopic (col. 6) and photometric (col. 7) classifications obtained in this work.

A complementary, albeit smaller, sample of 28 southern hemisphere eclipsing binaries is proposed by us for spectroscopic observations with the Fiber-fed Extended Range Optical Spectrograph \citep[{\sc feros};][]{Kaufer1999} instrument attached to the MPG/ESO 2.2-m telescope (under program 0106.A-0906(A)). Similar to the {\sc hermes}-based programme, we request eight epochs of spectroscopic observations and use exactly the same completion criteria for all the observed systems. The criteria are met for the total of 25 systems while the remaining 3 stars have received only 2 epochs of spectroscopic observations. Detailed information is provided in Table~\ref{Tab:targets} for the 25 systems that we consider completed, with the meaning of the columns being the same as for the {\sc hermes} sample.

The {\sc hermes} spectra are reduced with version 7.0 of the dedicated {\sc hermes} pipeline while the {\sc feros} spectra are processed with a modification of the {\sc ceres} pipeline \citep{Brahm2017} presented in \citet{Gebruers2022}. The data reduction steps are the standard ones and include bias subtraction, flat fielding, cosmic ray removal, wavelength calibration, barycentric correction, and order merging. Normalisation to the pseudo-continuum is done by selecting knot points and fitting a low-degree polynomial through them. Special care is taken in the regions of broad Balmer lines to ensure the outer line wings are not altered by the normalisation process.

\begin{figure*}
   \centering
   \includegraphics[width=18cm]{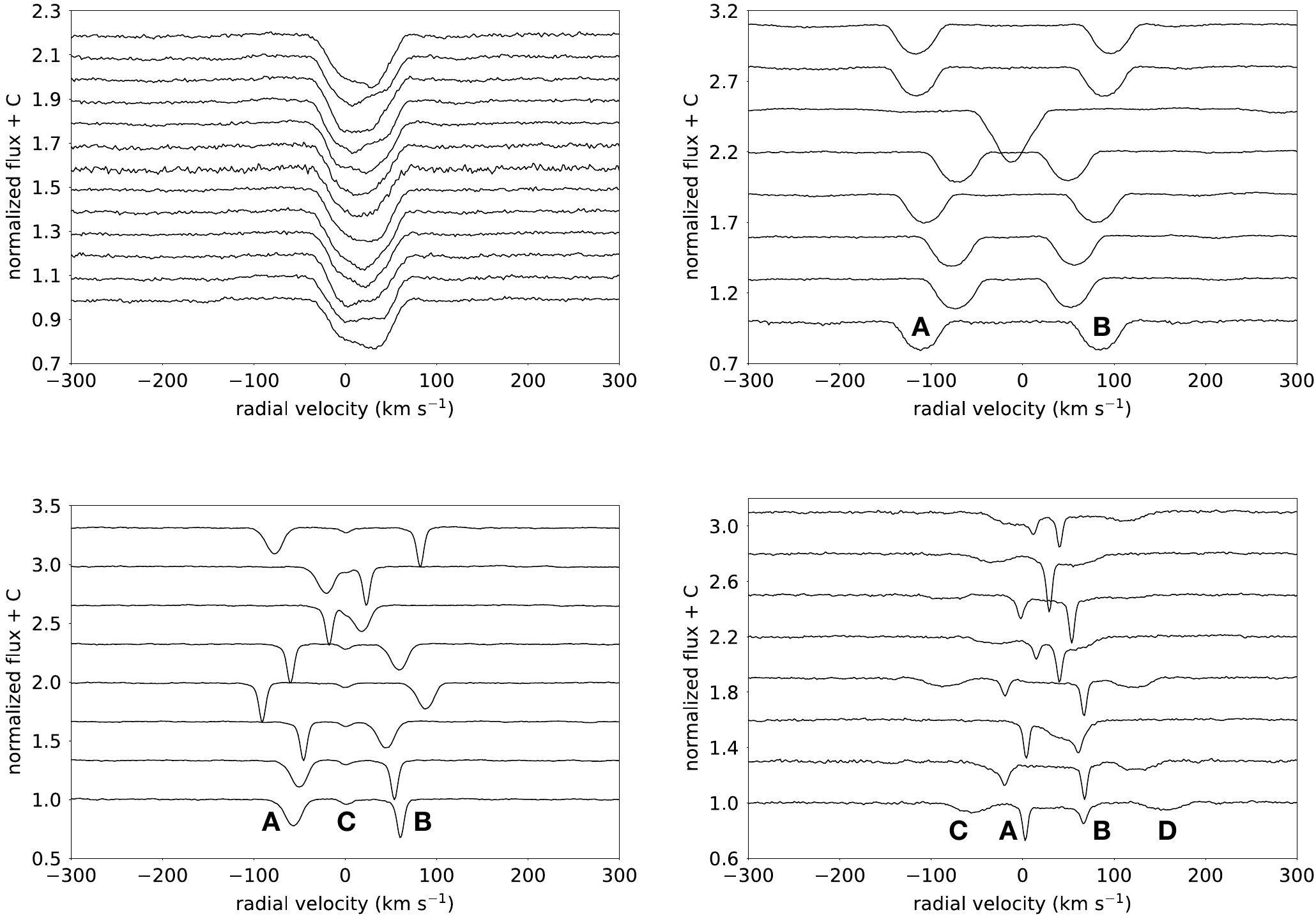}
      \caption{Examples of the LSD profiles computed with the \teff=11\,000~K mask. From top left to bottom right: HD~34382 - a single star showing line profile variability, HD~138305 - an SB2 binary system, HD~234650 - an SB3 triple system, and HD~57158 - an SB4 quadruple system. The individual LSD profiles in each panel have been shifted vertically by a constant factor for clarity.}
    \label{Fig:LSDProfiles}
\end{figure*}

\begin{figure*}
   \centering
\includegraphics[width=65mm]{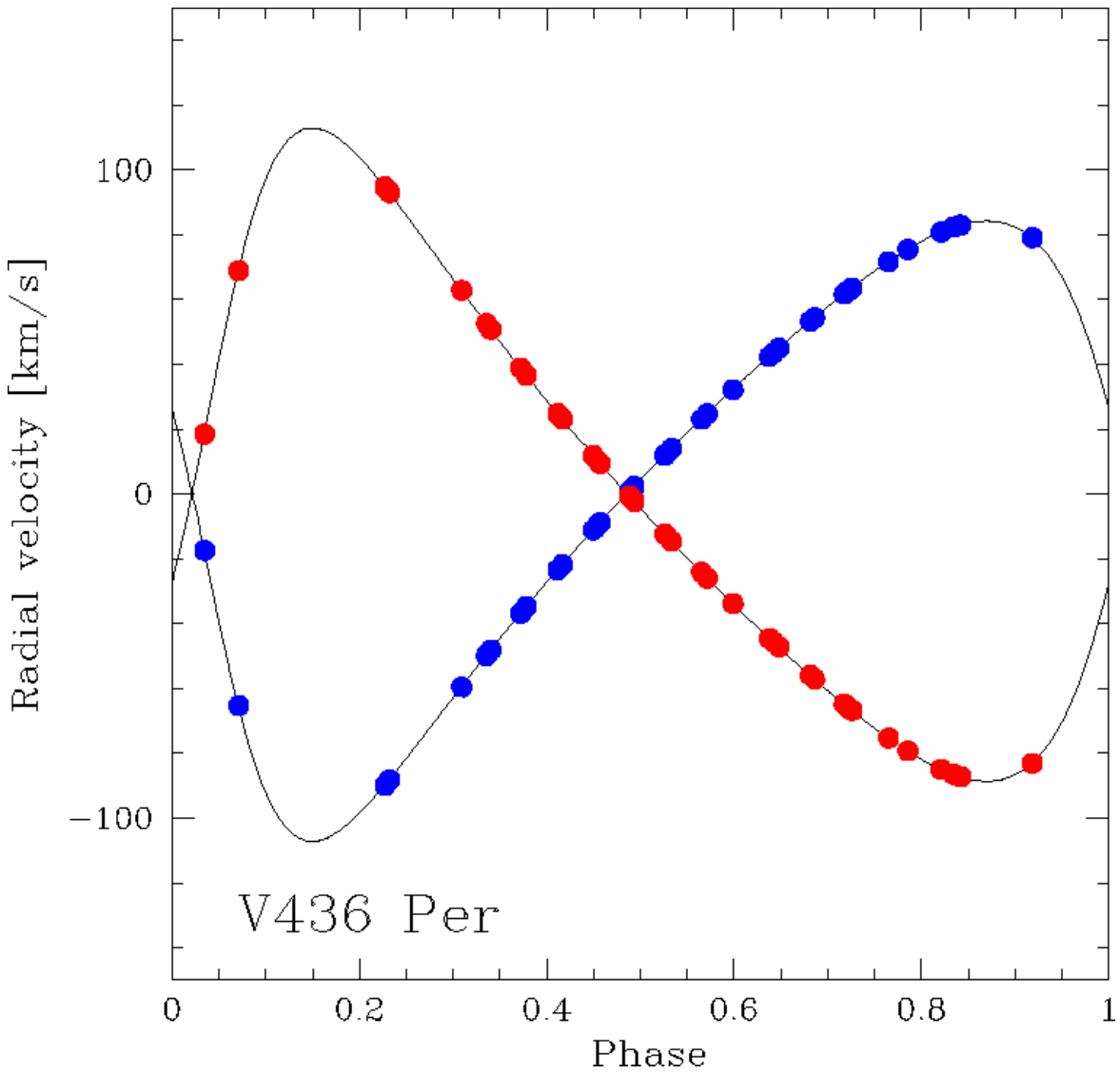}  \hspace{0.6cm}
\includegraphics[width=65mm]{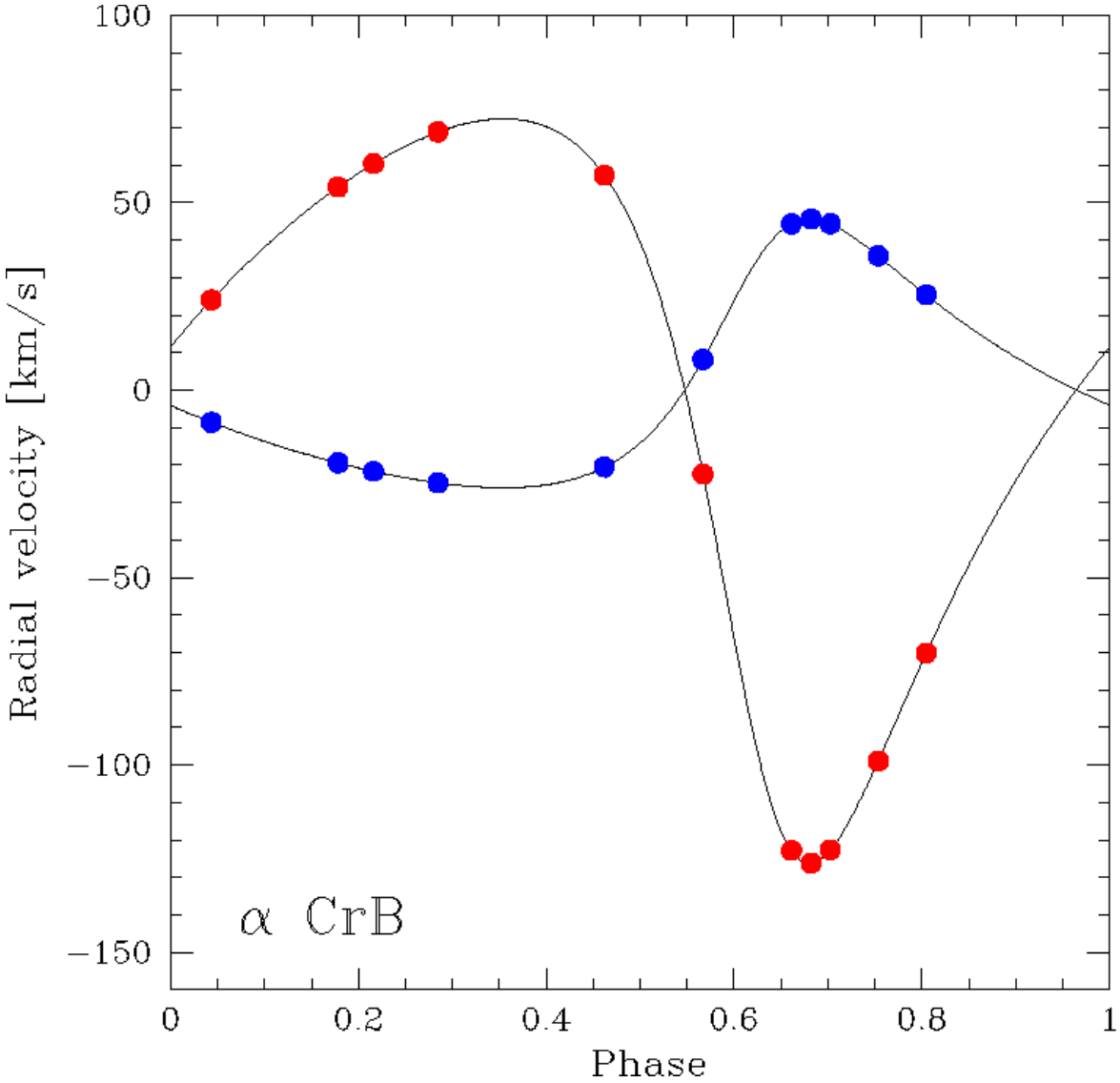} \\ \vspace{0.6cm} 
\includegraphics[width=65mm]{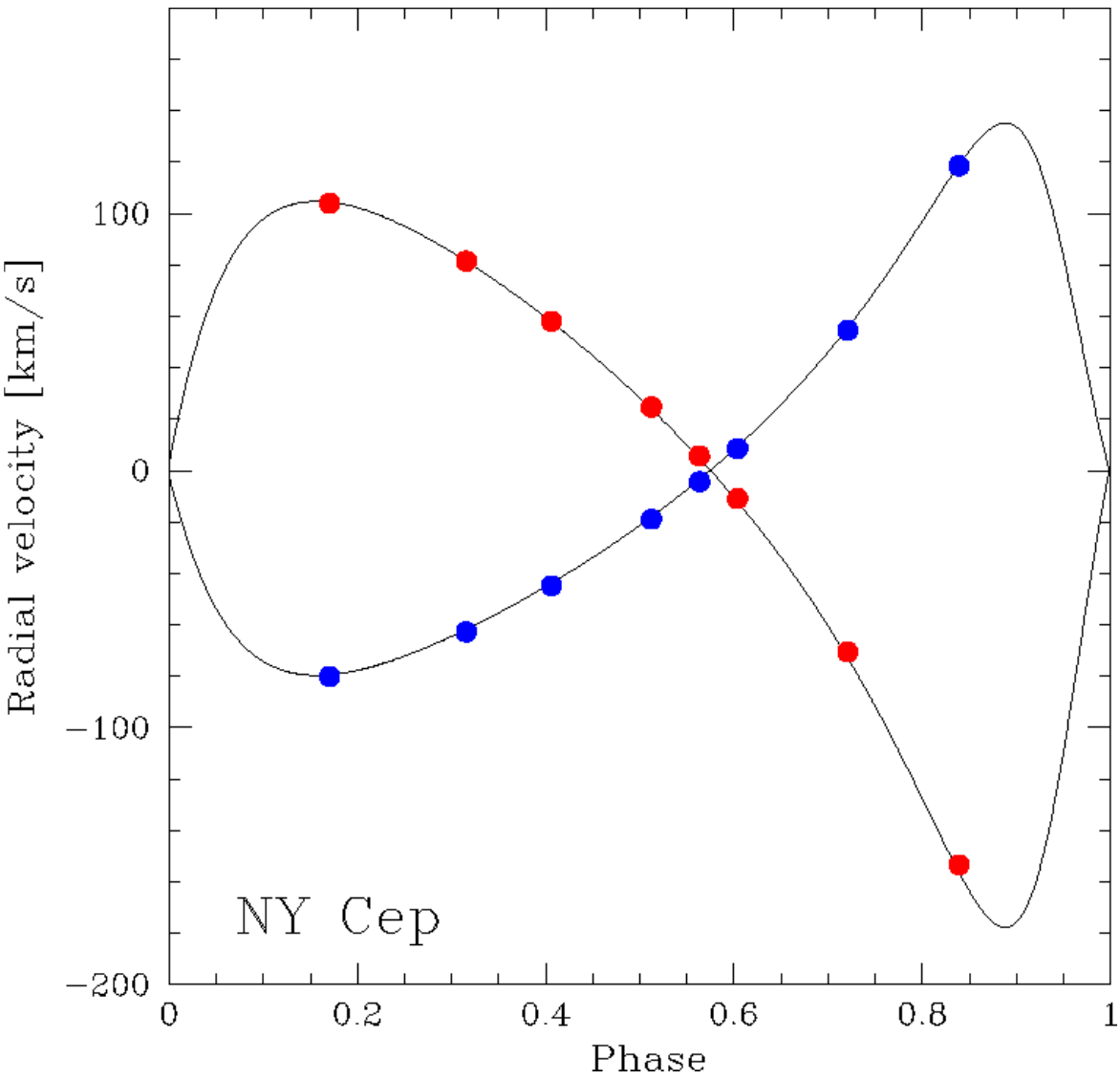}
\hspace{0.6cm} \includegraphics[width=65mm]{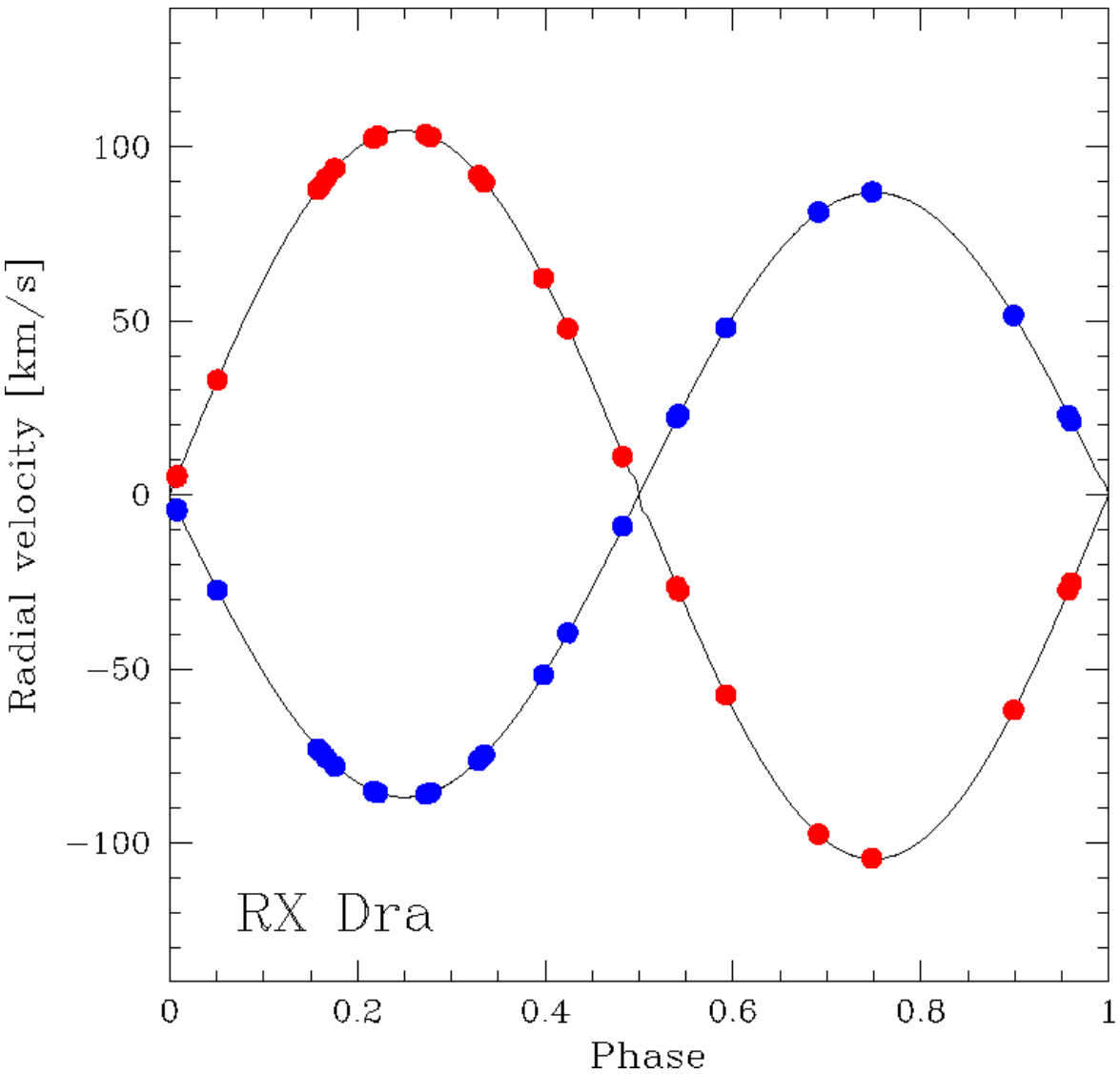} \\ \vspace{0.6cm} 
      \caption{Example orbital solutions for SB2 systems with differing numbers of observed spectra and fairly good phase coverage. Black lines depict the orbital solutions while blue and red dots represent orbital phases of the observed spectra for the primary and secondary components, respectively. V436~Per, $\alpha$~CrB and NY~Cep are highly eccentric binary systems with eccentricities of about 0.37, 0.38 and 0.44, respectively. RX~Dra represents the case of a circular-orbit system with the previously undetected secondary component that is revealed in this work.}
    \label{Fig:OrbitsExamples}
\end{figure*}

\begin{figure*}
   \centering
\includegraphics[width=80mm]{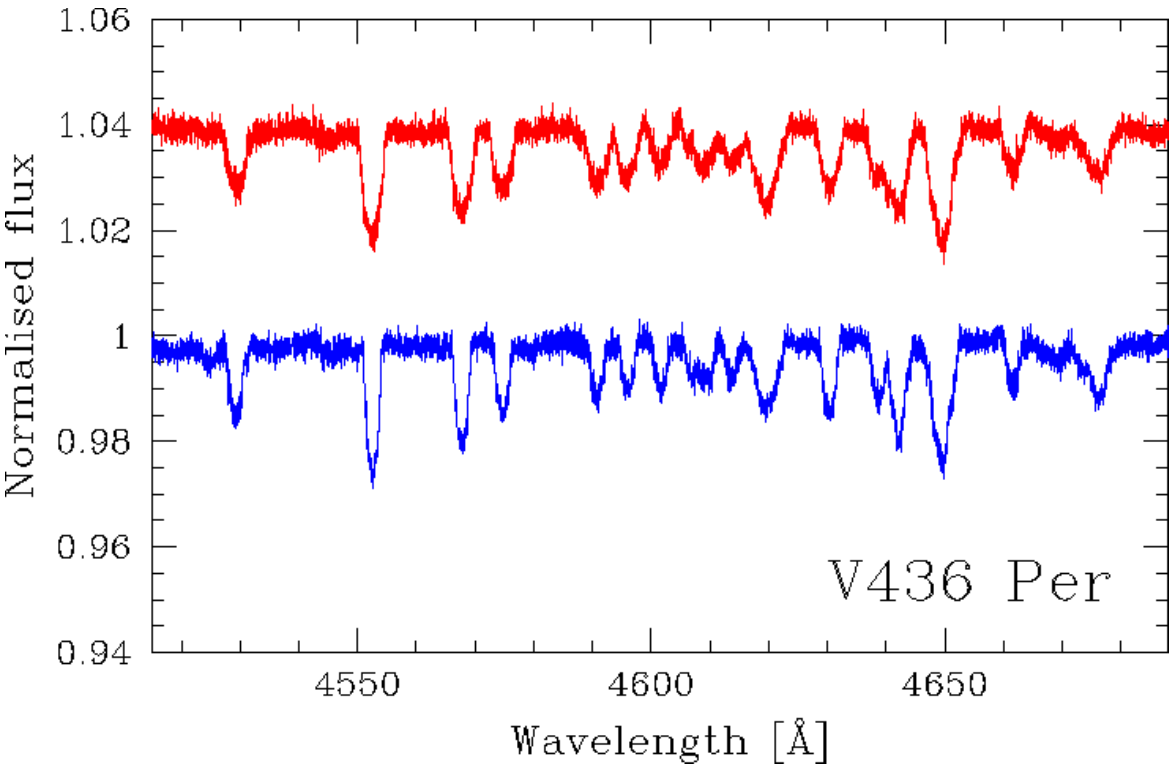}
 \hspace{0.5cm}   \includegraphics[width=80mm]{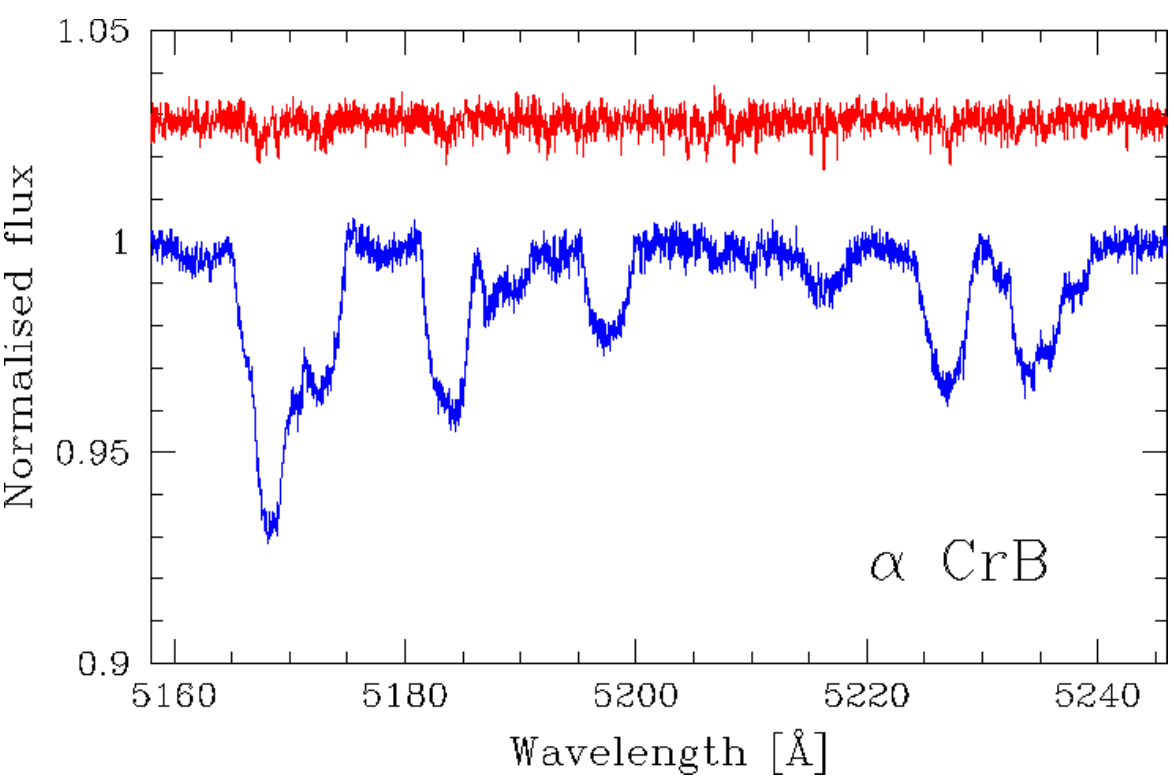} \\ \vspace{0.6cm}  \hspace{2mm}       \includegraphics[width=80mm]{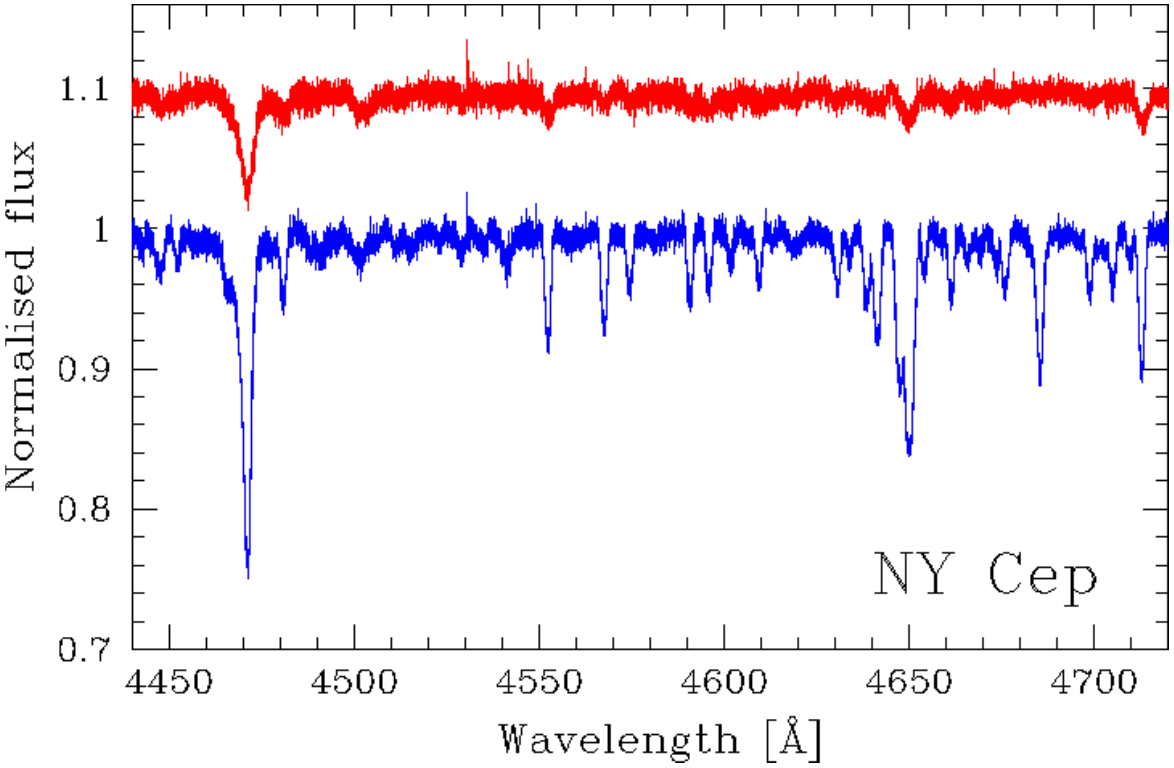}
\hspace{0.6cm}     \includegraphics[width=80mm]{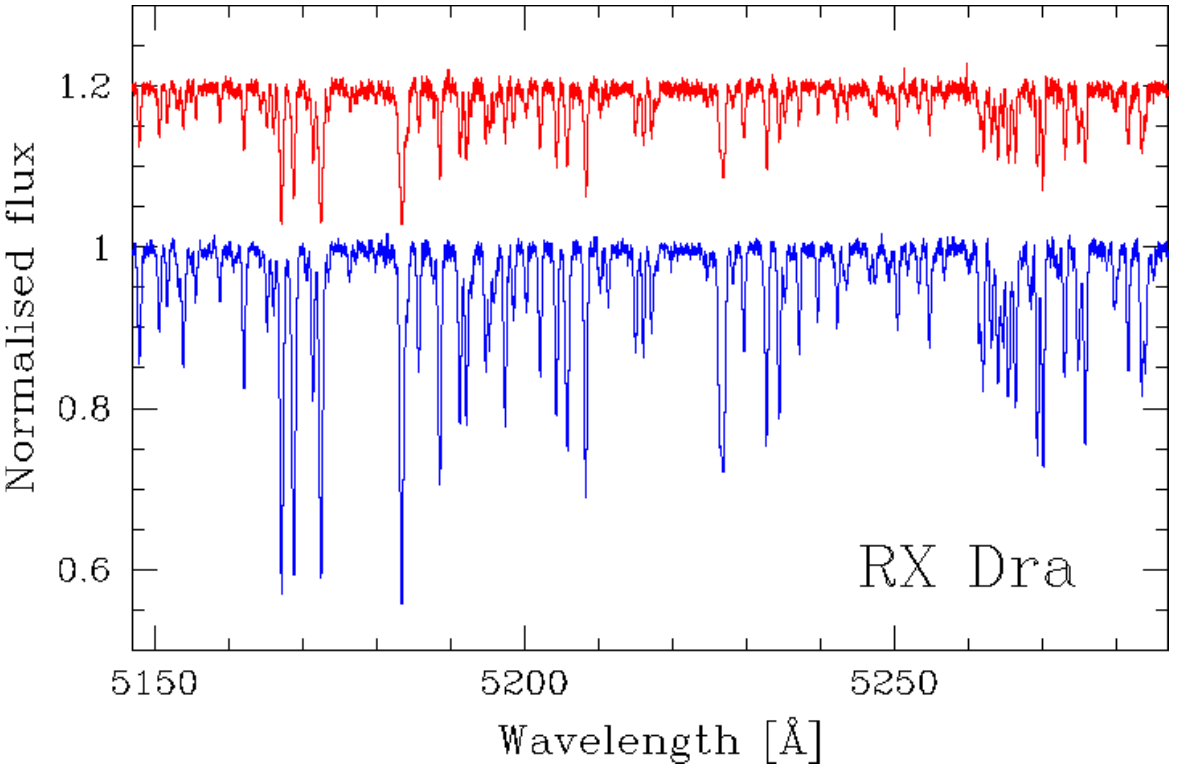} \\  \vspace{0.6cm}
      \caption{Examples of the disentangled spectra for the binary systems shown in Fig~\ref{Fig:OrbitsExamples}. In the upper row, V436~Per and $\alpha$~CrB represent systems with extreme light ratio. In the lower row, binary systems with hot (NY~Cep) and cool (RX Dra) components in our sample are shown for comparison.}
    \label{Fig:SPDspectra}
\end{figure*}

\section{Spectroscopic analysis}\label{sec: spectroscopic_analysis}

In this section, we present the results of our preliminary spectroscopic analysis of 83 systems, that is 58 in the {\sc hermes} sample and 25 in the {\sc feros} sample. We start with spectroscopic classification based on the least-squares deconvolution profiles (Section~\ref{sec:LSD}). The classification is validated and, if necessary, refined with the method of spectral disentangling which is subsequently used to compute the spectroscopic orbital elements of confirmed binary star systems (Section~\ref{sec:SPD}). 

\subsection{Least-squares deconvolution}\label{sec:LSD}

We employ the method of least-squares deconvolution \citep[LSD;][]{Donati1997} as implemented by \citet{Tkachenko2013}. The LSD method offers an efficient way to compute a mean profile from an ensemble of spectral lines present in the spectrum. Two fundamental assumptions of the method need to be kept in mind when computing an LSD profile: (i) all spectral lines should have similar shape, meaning that hydrogen, helium, and metal lines with damping wings need to be excluded from the calculations, and (ii) all spectral lines add up linearly implying that the depths of spectral blends formed of lines whose absorption coefficients overlap are not accurately reproduced by the method. Within these fundamental assumptions, stellar spectrum $I$ is represented as a convolution of a line mask $M$ with a priori unknown average profile $Z(v)$, such that
\begin{equation}
I = M \ast Z(v).
\end{equation}
The line mask consists of a set of delta functions that specify wavelength positions and theoretical strengths of spectral lines. This implies that the exact content of a line mask depends on the assumed atmospheric parameters (\teff, \logg, \vmicro, [M/H]) of the star, where the effective temperature has by far the most dominant effect. For the purpose of the analysis in this work, and informed by the spectral classification as OB(A)-type stars in \citet{IJspeert2021}, we compute line masks for two values of \teff, 8\,000~K and 11\,000~K, and use both of them for the calculation of the LSD profiles from every single spectrum of every object in the sample. The two masks are required and sufficient at the same time in order to account for the variable complexity of the stellar spectrum between these two temperature regimes.

We perform a visual classification of all systems based on the time series of their LSD profiles. Figure~\ref{Fig:LSDProfiles} shows time-series of the LSD profiles for four objects, ranging from a single star to a quadruple system. HD~34382 (top left) is classified by us as a single star exhibiting line profile variations (LPVs). The latter are caused by the intrinsic variability of the star either due to rotational modulation of inhomogeneities on its surface or stellar pulsations. HD~138305 (top right) is found by us to be an SB2 system composed of two similar stars. HD~234650 (bottom left) and HD~57158 (bottom right) are higher-order multiple systems with the former being a spectroscopic triple-lined (SB3) system while the latter being a spectroscopic quadruple-lined (SB4) system.

\subsection{Spectral disentangling}\label{sec:SPD}

The method of spectral disentangling (SPD) was originally introduced and formulated by \citet{Simon1994}. The method allows for a simultaneous, self-consistent inference of the orbital elements for spectroscopic binary and higher-order multiple systems and the disentangled spectra of the components these systems are composed of. SPD relies on the principle that the complex spectrum of a binary or multiple-star system is a linear combination of the individual components' spectra, each shifted due to the Doppler effect in the course of the orbital cycle and diluted by the component's fractional light contribution.

In the original formulation of \citet{Simon1994}, the disentangling problem is described by the following matrix equation:
\begin{equation}
\mathbf{M} \cdot \mathbf{x} = \mathbf{c},
\label{Eg:SPD_Matrix}
\end{equation}
where the single-column matrices $\mathbf{x}$ and $\mathbf{c}$ represent a priori unknown disentangled spectra of the components and the time-series of the observed spectra, respectively. The rectangular matrix $\mathbf{M}$ represents a linear transformation of $\mathbf{x}$ to $\mathbf{c}$, has a block structure, and is constructed using individual Doppler shifts for each of the components at each epoch of the observations. Matrix $\mathbf{M}$ also contains information on the light dilution factors for each component of the system and each observed spectrum in the time-series. In the case of high-resolution spectra and when the number of observed spectra exceeds the number of individual components, the system of linear equations is overdetermined (i.e. number of equations is greater than the number of unknowns). Mathematically, this is an ill-posed problem and regularisation conditions are required to solve it. \citet{Simon1994} employ an algebraic technique known as a singular value decomposition (SVD). Solving the system is typically a computationally demanding problem owing to the need to invert matrix $\mathbf{M}$  in Eq. (\ref{Eg:SPD_Matrix}).
 
Independently, a Fourier-based formulation of spectral disentangling is developed by \citet{Hadrava1995}. With the help of the the Fourier transform, a large set of equations can be uncoupled into many small sets of equations for each Fourier mode. Such an approach is significantly less computationally demanding than the original formulation. In either formulation, the SPD method can be exploited in two basic modes: spectrum disentangling, where a priori unknown components' spectra are optimised along with the orbital elements of the system, and spectrum separation, where RVs or orbital elements are assumed to be known and individual component's spectra are inferred from the respective fixed orbital solution \citep[e.g.][]{Pavlovski2010, Serenelli_2021}.

The SPD method builds on the following fundamental assumptions: (i) the dominant variability present in the observed composite spectra of a system should be due to the motion of the components around their common centre of mass; and (ii) observationally the orbital phase coverage should be complete and close to uniform to exploit the full power of the method. In this work, as well as in the future exploitation of our sample (cf. Section~\ref{sec:conclusions}), we are interested in the detection of high-contrast binary SB2 systems (i.e. where one of the binary components is significantly fainter than the other), including those where one of the components is a pulsating star. While the former represents a challenge for the method, the latter is, in principle, a direct violation of one of the method's fundamental assumption. Hence, we comment on both of these challenges in more detail to justify the choice of the SPD method for our purpose.

In the method of SPD, the gain in S/N in the final disentangled spectra is proportional to $\sqrt N$, where $N$ is the number of the observed spectra in a time-series. The gain is distributed according to the light ratio between the binary components. Previous studies demonstrate that SPD is a powerful tool to reveal the presence of a faint component (1--2\% contribution to the total light of the system) in high-contrast systems \citep[e.g.][]{Kolbas_2015, Themessl_2018, Pavlovski_2022, Johnston_2023}.

While time-dependent LPVs due to the effect of non-radial pulsations represent an extra source of variability that is not recognised by the SPD method, previous studies of binary systems with pulsating component(s) demonstrate that the method delivers reliable results provided that: (i) the LPV amplitude due to the intrinsic variability of the star is much smaller than due to its binary motion, and (ii) the orbital period of the system differs from that of the dominant intrinsic variability of the star \citep[e.g.][]{Uytterhoeven_2005a, Uytterhoeven_2005b, Ausseloos_2006, Tkachenko_2012}. For the case of a radially pulsating star where the RV amplitude of the dominant (radial) pulsation mode is comparable to or exceeds that of the orbital motion, a large set of observed spectra with uniform coverage of the pulsation and orbital cycles is required. Indeed, in such a scenario, a time series of the observed spectra can be binned with respect to the orbital phase of the binary system to average out the effect of oscillations on the line profile of a pulsating component. As a result, spectral line profiles in the orbital phase-binned observed spectra contain a minimum amount of distortion due to stellar pulsations, bringing us to the former case of orbital motion being the dominant source of the apparent line profile variability \citep[e.g.,][]{Tkachenko2014b, Tkachenko_2016}.

In this work, we employ the Fourier-based method of SPD as implemented in the {\sc FDBinary} software package\footnote{\url{http://sail.zpf.fer.hr/fdbinary/}} \citep{Ilijic2004}. The disentangling method is used by us to verify and if necessary refine the LSD-based spectroscopic classification and, more importantly, derive a preliminary set of orbital elements for all systems in our sample. The most common, yet not frequent, case of a refined classification is the one of an SB1 (LSD-based) to an SB2 (SPD-based) system. This is thanks to the power of the SPD method to detect spectral contributions as faint as a few percent (in the continuum flux units) which may stay unnoticed in the LSD profiles should not the most optimal line mask be chosen for the respective calculations.

The results of our combined LSD- and SPD-based spectroscopic classification are provided in Table~\ref{Tab:targets} (sixth column). For all SB1 and SB2 systems, we employ the {\sc FDBinary} code to compute their spectroscopic orbital elements, that is orbital eccentricity $e$, argument of periastron $\omega$, and semi-amplitudes of the individual components $K_{\rm i}$. Owing to the fact that the orbital period value can be constrained with typically a much higher precision from the eclipses in space-based photometric data than from a limited set of ground-based spectroscopic observations, we do not optimise the orbital period value in the SPD method by default. Instead, we improve the period separately using an algorithm described in detail in IJspeert et al. (in prep.) that searches globally for the best orbital period and then refines it with a granularity of one part in one hundred thousand. The global merit function is a combination of the phase dispersion measure \citep{Stellingwerf1978} and Lomb-Scargle amplitude \citep{Lomb1976,Scargle1982}, as well as the number and completeness factor of orbital harmonic frequencies found in the light curve. This period finding algorithm is part of a larger methodological framework for analysing eclipsing binary light curves developed by IJspeert et al. (in prep.) and implemented in STAR\_SHADOW\footnote{STAR\_SHADOW will be part of the mentioned publication and published on GitHub.} Furthermore, eccentricity is optimised by us in those case only where evidence for a value different from zero is found in the TESS light curve. Ultimately, the SPD solution is computed in three different wavelength regions free of Balmer lines and rich in metal lines, 4375-4490~\AA, 4510-4725~\AA, and 4900-5230~\AA. The orbital elements and their uncertainties are reported in Table~\ref{Tab:SPDOrbit} and are the mean and standard deviation of the mean values, respectively. We note that a typical S/N of about 80 or higher has been achieved in all spectroscopic epochs for all systems in the studied sample. This is sufficiently high to have a negligible effect on the estimated uncertainties of the obtained orbital elements compared to other contributors to the total error budget. These dominant contributors to the uncertainties are: (i) flux ratio of the binary components where lower flux contribution implies larger uncertainties on the K-value of the star and eccentricity of the system; (ii) spectral types of the binary components where a high density of lines in the spectra of cooler stars leads to more appreciable constraints on the K-values of stars and eccentricity of the orbit; and (iii) spectral line broadening with narrow spectral lines being more constraining on individual spectral shifts in the Fourier space and, as a result, on the eccentricity of the system and K-value of the respective binary component. While we opt for a simple method of uncertainty estimate in this work, our future studies will include a more sophisticated way of estimating uncertainties (e.g., the bootstrap method or jackknife estimator) that will allows us to account for the above-mentioned contributors to the total error budget.

Figure~\ref{Fig:OrbitsExamples} shows example orbital solutions for a selection of four systems from the studied sample. Black lines represent orbital solutions for both binary components as determined with the SPD method. Red and blue filled circles indicate, respectively, the predicted primary and secondary RVs from the obtained orbital solution and at the epochs of the acquired spectroscopic observations. We note that the method of SPD bypasses the step of RV determination from the observed spectra, hence the RVs shown in Figure~\ref{Fig:OrbitsExamples} are inferred quantities and used to illustrate typical phase coverage we achieve for our systems irrespective of their orbital eccentricity value. Examples of the disentangled spectra for the same four systems are shown in Figure~\ref{Fig:SPDspectra}. In the top row, we show a system with two similar components \citep[V436~Per, light ratio $l_{\rm B}/l_{\rm A} = 0.798\pm0.021$;][]{Southworth2022a} and a system with an extreme light ratio between the two stars ($\alpha$~CrB, light ratio $l_{\rm B}/l_{\rm A} = 0.0180\pm0.0002$, Pavlovski et al., in preparation). In the latter case, spectral lines of the secondary component have a depth of at most 0.5\% in the continuum units owing to their rotational broadening. In the bottom row, the disentangled spectra are shown for the NY~Cep (left) and RX~Dra (right) systems. The primary component in the detached eclipsing binary NY~Cep is one of the hottest stars in our sample which is evidenced by the presence of \ion{He}{ii} lines at 4541~{\AA} and 4687~{\AA}. SPD also reveals the spectrum of the secondary component in the RX~Dra system (right) which makes it an SB2 system with a $\gamma$~Dor pulsating component.

\section{Photometric analysis}\label{sec: photometric_analysis}

Previous efforts have demonstrated the power of combining high-quality TESS light curves with spectroscopic radial velocities for relatively small numbers of high-mass pulsating EBs (see e.g. \citealt{Bowman2019d, Southworth2020c, Southworth2021a, Southworth2022a}). As discussed in Section~\ref{sec:sample_selection}, our large sample of EBs was originally identified as eclipsing systems by \citet{IJspeert2021} using the products of automated light curve extraction tools, such as the MAST SPOC pipeline \citep{Jenkins2016b} and/or the TESS quick-look pipeline (QLP). However, in our current work we extract a light curve optimised for each pulsating EB using the TESS full-frame images (FFIs) following a similar methodology described in \citet{Bowman2022a}, which is briefly summarised here for completeness.

TESS pixel cutouts sized $25 \times 25$ are extracted from the FFIs using {\sc astrocut} \citep{Brasseur2019c}. We use the {\sc lightkurve} software package \citep{Lightkurve2018} to select aperture masks optimised to maximise the signal-to-noise of the eclipses (and pulsations if present) whilst minimising the contribution from nearby contaminating sources based on cross-checking with the location and fluxes of sources in the Gaia catalogue. The background flux is estimated from the median observed flux per frame after excluding pixels that contain flux from the target or other sources. The background flux is subtracted, and the extracted light curves are normalised by dividing through the median flux. Finally, we perform a principal component analysis (PCA) to remove remaining instrumental systematics in the light curves \citep{Bowman2022a}.

Due to the difference in integration times between different TESS data cycles and the large (i.e. $>1$~yr) gaps between cycles, we choose to analyse data from each cycle, if available, separately for each star. To classify the dominant variability beyond the eclipses caused by binarity, and ascertain if any additional signal is caused by pulsations or rotational modulation (RotMod) in our sample, we create a multi-frequency cosinusoid model to remove the orbital harmonics from the extracted TESS light. Our approach is similar to the method previously used by \citet{Bowman2019d} for the pulsating eclipsing binary system U~Gru. After optimising the orbital period and the number of significant orbital harmonic frequencies using a least-squares fit to the light curve, we subtract the resultant optimised analytical model from the light curve and calculate the residual amplitude spectrum \citep{Kurtz1985b}. The dominant variability is identified as pulsations or RotMod based on visual inspection of the residual amplitude spectrum. We apply this method to all the combined sectors within each TESS cycle to compare our classifications. For all stars, we find that our classification results are consistent across the multiple cycles, if available.

\begin{figure*}
    \centering
    \includegraphics[width=0.95\textwidth]{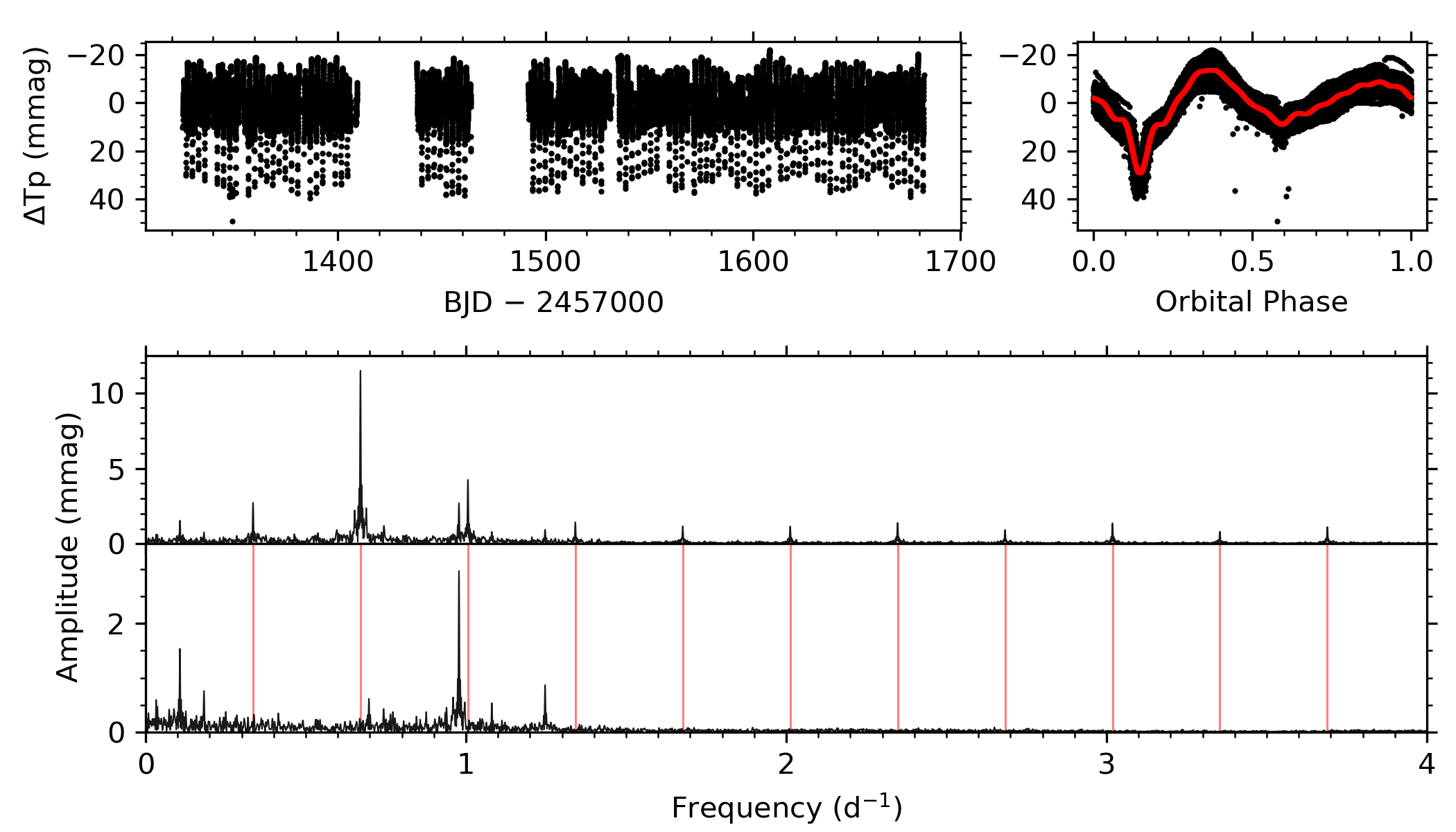}
    \caption{Example summary figure for TIC~150442264. Top panels: FFI-extracted light curve versus time (left) and phase-folded on the orbital period for all available cycle 1 sectors (right). Bottom panels: Original and residual amplitude spectra in which vertical red lines denote the location of significant orbital harmonics that comprise the analytic binary model, which is also shown as a red line in the phase-folded light curve panel. }
    \label{fig: TESS FT}
\end{figure*}

An example figure summarising this methodology is provided in Fig.~\ref{fig: TESS FT}, which contains our FFI-extracted light curve for all available cycle 1 data, the corresponding binary orbit phase-folded light curve, and the original and residual amplitude spectra for TIC~150442264 (HD~46792). In this example, several significant g-mode frequencies can be seen around 1~d$^{-1}$, and also a series of harmonics typical of rotational modulation at 0.1~d$^{-1}$. There are no significant pulsation frequencies above about 2~d$^{-1}$. We leave the combined binary modelling and pulsational frequency analysis for a future paper, but this approach allows us to classify the dominant pulsational variability in our targets and identify promising systems for future asteroseismic modelling, which are listed in Table~\ref{Tab:targets}. In so doing, we dramatically increase the number of pulsating EBs which span spectral types from O through to F, and include $\beta$~Cep through to $\gamma$~Dor pulsators.

\section{Discussion}\label{sec:discussion}
In this section, we present a summary of the obtained results and discuss them in the context of the previous studies. We recall that our sample selection was done in a blind way based on colour information and available TESS light curves. This approach to the target selection guarantees the presence in our sample of some of the well-studied systems in the literature allowing us to use those as the methodology validators.

We report the detection of 12 single stars that are found by us to show either LPVs or no variations at all. We note that two more systems are found at the interface of their (low-amplitude) LPVs being interpreted as due to binarity and intrinsic variability of the star. Fifteen stars are specrtoscopically classified by us as apparent SB1 systems. We note that the SB1 classification is not set in stone and may be revised into an SB2 when/if more spectra become available in the future. The method of spectral disentangling delivers average disentangled spectra with a gain in S/N of roughly a factor $\sqrt{N}$ with respect to the individual observed composite spectra, where $N$ stands for the number of spectra in a time-series. This implies that the detection limit of a hypothetical faint companion star with the SPD method increases with the number of observed spectra in a time-series which may ultimately lead to reclassification of a star from SB1 into SB2. Furthermore, 50 targets are classified by us as SB2 systems. These are the primary candidates for future studies in the context of the observational mapping of the mass discrepancy. Finally, we report the detection of four high-order multiple systems, of which two objects are classified by us as triple (SB3) and two as quadruple (SB4) systems.

Twenty eclipsing systems are classified by us as presumably containing at least one oscillating component. Of these twenty systems, one (u Tau) is a tentative detection where we cannot exclude rotational modulation as a cause of the intrinsic variability of the star, and two systems (V436 Per and HD 84493) are likely to contain a component showing stochastic low-frequency variability \citep[e.g.,][]{Bowman2019,Bowman2020a}. Sixteen systems are found to contain either a g- (eight binaries) or p-mode (eight binaries) pulsator, and one system (V350 And) contains a g-/p-mode hybrid pulsator. Finally, the CD-45 4393 system is found by us to show a steadily decreasing eclipse depth across different TESS sectors such that the signal practically disappears in sectors 61 and 62. Such a phenomenon can be a manifestation of the precession of the binary orbit (e.g. due to the presence of a third body in the system) that causes the orbital inclination angle with respect to the observer's line-of-sight to change with time.  

Below, we discuss systems that have been studied in the literature and, where applicable, compare our photometric and spectroscopic results with those reported by other research groups. In doing that, we focus primarily on dedicated studies rather than on large catalogues, hence the overview provided below is probably not inclusive.

BD+13 1880: \citet{Subjak2020} report the system to comprise a metallic-line Am star and a brown dwarf companion in a 3.6772$\pm$0.0001~d orbit. From the analysis of a set of some 50 spectra obtained with multiple high-resolution spectrographs, the authors report the RV semi-amplitude $K$ of the host star of 4.64$\pm$0.03~\kms, in a good agreement with the value of 5.0$\pm$0.1~\kms\ derived in this work from a time-series of 10 {\sc hermes} spectra.

BD+36 3317: \citet{Ozdarcan2012} classify the system as an Algol-type binary based on the analysis of ground-based multi-colour photometry, and provide evidence of the system being a member of the $\delta$ Lyrae cluster. The system is revisited by \citet{Kiran2016} based on photometric observations from the literature and a newly obtained time-series of 20 medium-resolution ($R=11\,700$) spectra. The authors reinforce the conclusion on the cluster membership of the system, and report the RV semi-amplitudes of the primary and secondary to be $K_{\rm 1}$=80.8$\pm$1.1 and $K_{\rm 2}$=124.1$\pm$1.3, respectively. Both values are in good agreement with those of $K_{\rm 1}$=82.93$\pm$0.03 and $K_{\rm 2}$=127.3$\pm$0.1 derived in this work from 8 {\sc hermes} spectra.

65~UMa~B: the star is discovered as magnetic by \citet{Bychkov2003} while \citet{Auriere2007} measure its longitudinal magnetic field and rotation period to be $B_l=-166\pm 20$~G and 15.830~d, respectively. \citet{Joshi2010} present a detailed spectroscopic study of the star based on a time-series of high-resolution spectra obtained with the 2.56-m Nordic Optical Telescope (NOT). The authors report on the absence of RV variations in the spectra and on the abundance pattern characteristic of evolved Ap stars. \citet[][see their fig.~7]{Zasche2012} report 65~UMa to be a sextuple system consisting of: (i) a close eclipsing pair of nearly identical stars (Aa1+Aa2) orbiting each other with a period of $\sim$1.73~d; (ii) a distant third component (Ab) orbiting the close pair with a period of about 640~d; (iii) a fourth component (65~UMa~B) resolved interferometrically whose period is about 118~yr; and (iv) fifth (65~UMa~C) and sixth (65~UMa~D) components with periods of $\sim$14~kyr and $\sim$591~kyr, respectively. Our spectroscopic classification of 65 UMa B as a single star not showing notable RV variability is in full agreement with the literature. The TESS light curve reveals the presence of shallow eclipses indicative of an eclipsing binary with a period of about 1.73~d. We conclude that the eclipse signal in the TESS light curve of 65~UMa~B is due to a contaminating light from the component A which has a similar surface brightness.

EL CVn: \citet{Maxted2014} study the system based on a combined WASP photometry and {\sc ses} high-resolution (R=60\,000) spectroscopy. The authors find the system to be composed of an A-type primary and a pre-He white dwarf companion, residing in a circular orbit with the RV semi-amplitude of the primary $K_{\rm 1}=29.0\pm 0.4$~\kms. The latter is in agreement with our determination of $K_{\rm 1}=30.0\pm 1.6$~\kms. \citet{Wang2020} measure the spectral lines of both binary components with the far-UV HST/COS spectroscopy and report the detection of a spectral signature of the pre-He-WD companion in the \ion{Mg}{ii}~4481~\AA\ optical line in the medium-resolution (R=31\,500) {\sc arces} spectra. The authors report $K_{\rm 1}=29.0\pm 0.4$~\kms\ and $K_{\rm 2}=236.2\pm 1.1$~\kms, in good agreement with the determination of the respective parameters in this work. 

$\alpha$~CrB is a well-studied object whose orbit is reported for the first time by \citet{Jordan1910}. The system is solved spectroscopically by \citet{Ebbighausen1976} who reports $e=0.404\pm 0.004$ and $K_{\rm 1}=35.8\pm 0.2$~\kms. \citet{Tomkin1986} report the detection of a secondary component based on newly obtained spectroscopic data with the coud\'{e} spectrograph attached to the 2.7-m McDonald Observatory telescope. The authors report $e=0.371\pm 0.005$, $K_{\rm 1}=35.4\pm 0.5$~\kms, and $K_{\rm 2}=99.0\pm 0.5$~\kms, in good agreement with the respective parameters derived in this work from 14 {\sc hermes} spectra. Of the more recent studies, \citet{Schmitt2016} provide RV measurements of the fainter secondary component using the method of cross-correlation and combine those with their own and historical RV data of the primary component to update the system's orbital elements and to search for evidence of an apsidal motion. The authors report $e=0.379\pm 0.002$, $K_{\rm 1}=36.2\pm 0.1$~\kms, and $K_{\rm 2}=98.0\pm 0.3$~\kms, in excellent agreement with the respective orbital elements derived in this work. In addition, \citet{Schmitt2016} find the apsidal motion period to be between 6600 and 10\,600 years in agreement with the previous findings, and report on the alignment of the orbit and rotation axes.

V994 Her: the system is discovered to have a quadruple-lined double-eclipsing nature from ground-based multi-colour photometric and high-resolution spectroscopic data by \citet{Lee2008}. The authors report two eclipsing binaries to have orbital periods of $\sim$2.08~d and $\sim$1.42~d, of which the former is in agreement with the period used in this work and reported by \citet{IJspeert2021}. Both binaries are found to have slightly elliptic ($e$<0.1) orbits with the reported masses of 2.83$\pm$0.20~M$_{\odot}$, 2.30$\pm$0.16~M$_{\odot}$, 1.87$\pm$0.12~M$_{\odot}$, and 1.86$\pm$0.12~M$_{\odot}$ for the Aa, Ab, Ba, and Bb components, respectively. \citet{Zasche2013} report V994 Her to be a quintuple system with the two eclipsing binaries orbiting each other with the period of about 6.3~d. These findings are refined in \citet{Zasche2016} where the authors report orbital period of $\sim$2.9~d for the two eclipsing pairs and find evidence for apsidal motion with periods of about 116 and 111 years. The authors also report updated masses of 3.01$\pm$0.06~M$_{\odot}$, 2.58$\pm$0.05~M$_{\odot}$, 1.84$\pm$0.03~M$_{\odot}$, and 1.93$\pm$0.04~M$_{\odot}$ for the Aa, Ab, Ba, and Bb components, respectively. Ultimately, \citet{Zasche2023} report the discovery of a third set of eclipses in the TESS space-based and archival ground-based photometric data which makes the system a triply-eclipsing sextuple star system. The authors refine the orbital period of the A(Aa+Ab)--B(Ba+Bb) core system to 1062$\pm$2~d and report about its close to 3:2 mean motion resonance. The eclipsing pair C(Ca+Cb) is reported to have an elliptic orbit with the period of 1.96~d and component masses of 1.81$^{+0.17}_{-0.07}$~M$_{\odot}$ and 1.08$^{+0.16}_{-0.11}$~M$_{\odot}$, respectively. Our classification of V994 Her as an SB4 system in this work is in agreement with the previous findings.

V1898 Cyg: the system is classified as a spectroscopic double-lined binary by \citet{Abt1972} and as an eclipsing system by \citet{Halbedel1985}. A detailed analysis of the system based on the newly obtained spectroscopic and archival photometric data is presented in \citet{Dervisoglu2011}. The authors report the $K_1$ and $K_2$ RV semi-amplitudes of 55.2$\pm$0.8~\kms\ and 287.6$\pm$2.1~\kms, respectively, in good agreement with the findings in this work.

GK Cep: the system is classified as an eclipsing spectroscopic double-lined binary with the period of 0.936171~d and spectroscopic mass ratio of 0.92 by \citet{Bartolini1965}. Furthermore, \citet{Pribulla2009} classify the system as a spectroscopic triple, in agreement with the findings in this work. A detailed study of GK Cep is presented in \citet{Zhao2021} based on newly obtained photometric data with the lunar-based ultraviolet telescope and TESS space-based mission. The authors confirm the presence of a third body in the system and measure the masses of the close binary components to be 1.93~M$_{\odot}$ and 2.11~M$_{\odot}$. 

AH Cep: the system is analysed based on a combined photometric and spectroscopic data set by \citet{Bell1986}. The authors report $K_1$ = 249$\pm$8~\kms\ and $K_2$ = 283$\pm$8~\kms\ under the assumption of a circular orbit. \citet{Holmgren1990a} and \citet{Burkholder1997} report $K_1$ = 237$\pm$2~\kms\ and $K_2$ = 269$\pm$2~\kms\ and $K_1$ = 230.0$\pm$3.2~\kms\ and $K_2$ = 277.6$\pm$4.4~\kms, respectively, under the same assumption of $e=0$. The latter solution is in agreement with the findings in this work. A recent study by \citet{Pavlovski2018} reports a slightly higher value for $K_1$ = 234.9$\pm$1.1~\kms\ while $K_2$ = 276.9$\pm$1.4~\kms\ remains to be consistent with our determination of the respective parameter.

WW Aur: a detailed analysis of the system is presented by \citet{Southworth2005}. The authors report $K_1$ = 116.81$\pm$0.23~\kms\ and $K_2$ = 126.49$\pm$0.28~\kms\ and $M_1$ = 1.964$\pm$0.007~M$_{\odot}$ and $M_2$ = 1.814$\pm$0.007~M$_{\odot}$ for the primary and secondary component, respectively. Our determinations of the RV semi-amplitudes are in agreement for the primary component and slightly lower for the secondary star.

AW Cam: a single-lined binary solution is presented by \citet{Mammano1967} with $K_1$ = 112$\pm$5~\kms\ under the assumption of a circular orbit. \citet{Frey2010} provide a simultaneous light- and RV-curve solution estimating $K_1$ to 110~\kms. Our estimate of $K_1$ is in agreement with the previous findings and we establish the double-lined nature of the system in this work.

V453 Cyg: \citet{Popper1991}, \citet{Simon1994}, and \citet{Burkholder1997} report spectroscopic RV semi-amplitudes of $K_1$ = 171$\pm$1.5~\kms, $K_2$ = 222$\pm$2.5~\kms; $K_1$ = 171.7$\pm$2.9~\kms, $K_2$ = 223.1$\pm$2.9~\kms; $K_1$ = 173.2$\pm$1.3~\kms, $K_2$ = 213.6$\pm$3.0~\kms, respectively. Of the more recent studies, \citet{Southworth2004} obtain eccentricity $e=0.022\pm 0.002$ from the apsidal motion solution and $K_1$ = 173.7$\pm$0.8~\kms, $K_2$ = 224.6$\pm$2.0~\kms\ from the RV fitting. Consistent with these findings are the solutions obtained by \citet{Pavlovski2009a,Pavlovski2018}, where the latter study reports $e=0.022\pm 0.002$, $K_1$ = 175.2$\pm$0.7~\kms, and $K_2$ = 220.2$\pm$1.6~\kms. Orbital elements determined by us in this work are in agreement with the previous studies, including small eccentricity of the system. The star is classified as including a $\beta$~Cep variable by \citet{Southworth2020c}, in line with our own classification in this work.

NY Cep: \citet{Holmgren1990b} present a detailed spectroscopic study of the system based on 26 newly obtained high-resolution spectra. Among other things, the authors report eccentricity, argument of periastron, and components' RV semi-amplitudes of $e=0.48\pm0.02$, $\omega = 58^{\circ}\pm2$, $K_1$ = 112$\pm$3~\kms, and $K_2$ = 158$\pm$8~\kms, respectively. \citet{Albrecht2011} revisit the system based on 46 newly obtained high-resolution spectra with the {\sc sophie} spectrograph. The authors report $e=0.443\pm0.005$, $\omega = 56.3^{\circ}\pm1$, $K_1$ = 113.8$\pm$1.2~\kms, and $K_2$ = 139$\pm$4~\kms\ assuming $P=15.27566$~d as derived in \citet{Ahn1992}. Our determinations of the eccentricity and argument of periastron of the system, as well as the RV semi-amplitudes of both binary components are in good agreement with the findings by \citet{Albrecht2011}.

16 Lac ($=$ EN Lac): a detailed spectroscopic analysis of the star as an SB1 system based on some 1200 newly obtained and archival spectra is presented by \citet{Lehmann2001}. The authors report $P=12.096844$~d, $e=0.0392\pm0.0017$, $\omega=63.7^{\circ}\pm2.1$, and $K_1$ = 23.818$\pm$0.033~\kms. While the system has also been known as a single eclipsing since the discovery by \citet{Jerzykiewicz1980}, a grazing secondary eclipse is for the first time detected in the TESS data by \citet{Southworth2022a}. The spectroscopic orbital elements derived by us in this work are in good agreement with those presented by \citet{Lehmann2001}. The star is reported as a $\beta$~Cep variable by \citet{Jerzykiewicz1980,Dziembowski1996,Lehmann2001,Aerts2003,Jerzykiewicz2015,Southworth2022a}, in line with our own findings in this work.

V436 Per: \citet{Harmanec1997} present the first detailed spectroscopic study of the system based on a collection of newly obtained and archival data. Using the method of spectral disentangling, the authors report the detection of LPVs and determine orbital eccentricity, argument of periastron, and RV semi-amplitudes of the components to be $e=0.3882\pm0.0043$, $\omega = 108.98^{\circ}\pm0.27$, $K_1$ = 98.0$\pm$1.0~\kms, and $K_2$ = 102.5$\pm$1.2~\kms, respectively. The system is revisited by \citet{Janik2003} based on a new set of high-resolution spectroscopic observations. The authors do not confirm the previously reported LPVs and present an updated set of spectroscopic orbital elements, $e=0.3768\pm0.0014$, $\omega = 109.83^{\circ}\pm0.10$, $K_1$ = 97.4$\pm$0.1~\kms, and $K_2$ = 91.2$\pm$0.1~\kms. A notable finding in the latter study is that the secondary component has a lower RV semi-amplitude suggesting it is a more massive star in the system. \citet{Southworth2022a} analyse TESS photometric data of the V346 Per system, report several local minima in the obtained solution, and emphasise the importance of obtaining an independent (spectroscopic) estimate of the components' light ratio to resolve the faced degeneracy. Spectroscopic orbital elements of the system derived by us in this work are in agreement with those reported by \citet{Harmanec1997} except that both $K$-values found by us are some 0.3\% lower. We also note that unlike \citet{Janik2003} we find the primary component to be the more massive one in the system. In addition, we find the star to exhibit stochastic low-frequency photometric variability, in a agreement with the classification by \citet{Southworth2022a}.

RX Dra: we find the system to contain a $\gamma$~Dor-type pulsator, in agreement with the classification by \citet{Southworth2022b}.

V1425 Cyg: a detailed study of the system based on a combined set of spectroscopic and photometric observations is presented by \citet{Hill1993}. The authors report $K_1$ = 142.3$\pm$1.8~\kms\ and $K_2$ = 221.7$\pm$2.2~\kms\ under the assumption of a circular orbit. The respective RV semi-amplitudes derived by us in this work are in agreement with the findings by \citet{Hill1993}.

V446 Cep: our photometric classification of the system as a $\beta$~Cep pulsator is in agreement with the conclusions by \citet{Southworth2022a} who also report the star to presumably exhibit tidally-induced oscillations.

CM Lac: a study of the system based on a combined set of spectroscopic and photometric observations is presented by \citet{Liakos2012}. The authors determine $K_1$ = 119$\pm$2~\kms, and $K_2$ = 156$\pm$2~\kms\ under the assumption of a circular orbit. \citet{Southworth2022b} revisit the system based on TESS photometric data and spectroscopic RV measurements from \citet{Liakos2012}. The authors report $K_1$ = 120.0$\pm$3.4~\kms, and $K_2$ = 157.0$\pm$3.3~\kms, in agreement with the previous findings. Our determinations of the RV semi-amplitudes are in good agreement with both of the above-mentioned studies. \citet{Southworth2022b} report the system to contain a $\gamma$~Dor pulsator, we confirm their findings in this work.

V398 Lac: the system is investigated spectroscopically by \citet{Cakirli2007}. The authors report $e=0.230$, $K_1$ = 110.3$\pm$3.7~\kms, and $K_2$ = 128.5$\pm$3.8~\kms\ for the eccentricity and RV semi-amplitudes of the components, respectively. A moderate orbital eccentricity of 0.273 and $0.2284\pm0.0007$ is confirmed by \citet{Bulut2008} and \citet{Wolf2013}, respectively, based on the studies of Hipparcos photometry and apsidal motion of the system. While the orbital eccentricity obtained by us is in agreement with the previous studies the RV semi-amplitudes of both components are lower by some 15\%-20\% than those reported by \citet{Cakirli2007}. 

V402 Lac: a detailed analysis of the system based on the newly obtained spectroscopic data and TESS space-based photometry is presented by \citet{Baroch2022}. The authors determine eccentricity and RV semi-amplitudes to be $e=0.376\pm0.003$, $K_1$ = 128.5$\pm$0.8~\kms, and $K_2$ = 129.2$\pm$0.8~\kms\ from the combined RV and eclipse timing analysis, suggesting the system is composed of two stars of a similar mass. Our findings are similar to but not exactly the same as the determinations by \citet{Baroch2022}. In particular, we find a slightly lower eccentricity and a $K$-value for the primary component, the latter being suggestive of a slightly smaller mass ratio of the system.

AE Pic: spectroscopic orbit of the system is presented by \citet{Sahade1950}. The authors report $e=0.1$, $\omega = 39^{\circ}$, and $K_1$ = 119~\kms. In this work, we classify the star as an SB2 system with a slightly lower eccentricity and RV semi-amplitude of the primary than reported previously. The system is classified as an eclipsing binary containing a pulsator that additionally exhibits signatures of rotational modulation by \citet{Barraza2022}. We confirm their findings in this work and classify the star as to having a $\beta$~Cep pulsator with extra signatures of rotational modulation in the light curve.

\section{Conclusions}\label{sec:conclusions}

\begin{figure}
   \includegraphics[width=9.8cm]{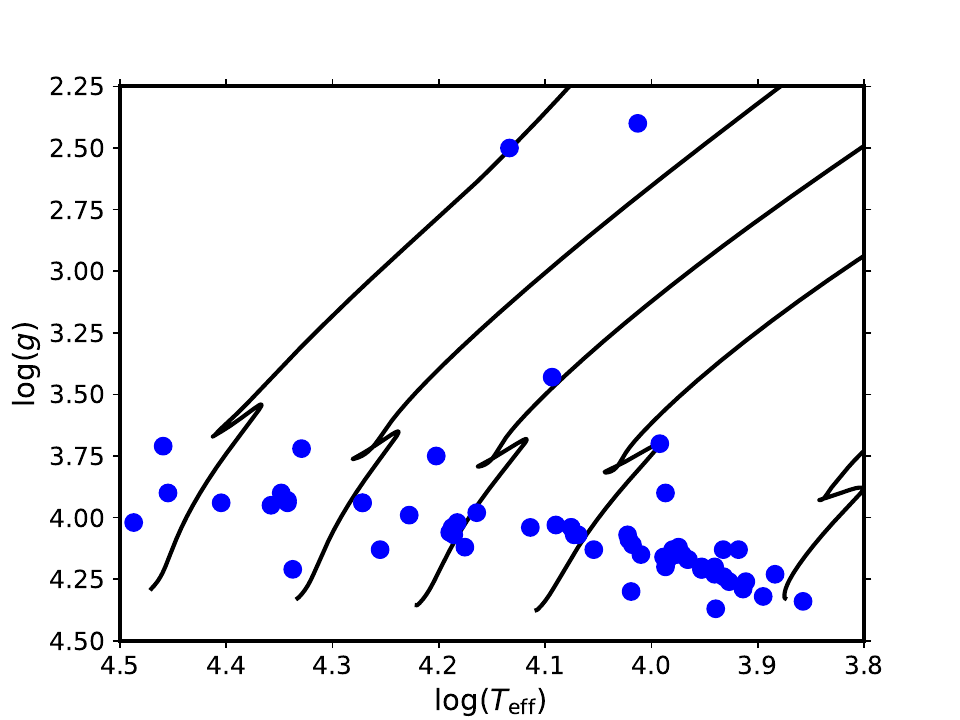}
      \caption{Kiel diagram of stars listed in Table~\ref{Tab:SPDOrbit}. In case of SB2 systems, parameters of the hotter component are displayed. Solid lines represent {\sc mesa} evolutionary tracks for the stellar mass values $M$ of, from right to left, 1.5, 3.0, 4.5, 7,0, and 13 M$_{\odot}$, and a minimum amount of overshoot $f_{\rm OV}=0.005$~$H_{\rm p}$. The tracks are from \citet{Johnston2019a}.}
    \label{Fig:KielDiagram}
\end{figure}

Of the total of 573 systems proposed for spectroscopic monitoring with the {\sc hermes} and {\sc feros} spectrographs, in this work we report phase-resolved spectroscopic observations for 83 of them (cf.\ Table~\ref{Tab:targets}). Sixty-five systems are classified by us as either SB1 or SB2 and their respective orbital elements obtained with the method of spectral disentangling are provided in Table~\ref{Tab:SPDOrbit}. For two more systems, HD~246047 and CD-47~4364, neither the method of spectral disentangling nor the LSD-based classification could give a definitive answer as to whether these stars are single and show low-amplitude LPVs or rather single-lined binaries with low $K$-values. Furthermore, we identify in total four high-order SB3 or SB4 multiple systems, while the remaining 12 targets are classified by us as single stars.

The $\log(T_{\rm eff})-\log(g)$ Kiel diagram of the above-mentioned sixty five SB1 and SB2 systems is shown in Figure~\ref{Fig:KielDiagram}. Where possible, the \teff\ and \logg\ values have been taken from literature (cf. Section~\ref{sec:discussion}), otherwise the respective values have been estimated from the spectral type and luminosity class of the star by interpolating in the tables of \citet{Schmidt-Kaler1982}. We also note that \teff\ and \logg\ of only the hotter and more massive primary components are displayed in Fig.~\ref{Fig:KielDiagram}. Therefore, the positions of the majority of stars in the Kiel diagram are not precise but rather indicative and are used for the purpose of demonstration of the parameter space covered by the sample presented in this work. With a slightly over 10\% sub-sample of the total 573 systems that we are currently monitoring spectroscopically, we achieve the coverage of a large range in the effective temperature and mass of the star, i.e. \teff\ $\in [7000,30000]$~K and $M\in [1.5,15]$~M$_{\odot}$. Currently, the lower-mass regime with $M\lesssim 3.5$~M$_{\odot}$ is populated more densely and the sample consists almost exclusively of stars in the core hydrogen burning phase.

Nevertheless, the present sample allows us to start looking into the problem of the mass discrepancy as a function of the stellar mass, surface properties of the star (i.e. \teff, \logg, and $v\,\sin\,i$), binary configuration (i.e. detached versus semi-detached systems), and intrinsic variability of the star (i.e. pulsating versus non-pulsating components). Indeed, several systems in the sample with AF-type primary components are known Algols and more of those have light curve characteristics of Algol-type binaries. A mix of well-detached and semi-detached systems in the sample allows us to assess quantitatively the role of binary interactions in the mass discrepancy problem, as has been suggested by \citet{Mahy2020}. Furthermore, diversity of the sample in terms of the effective temperature and mass of the primary component is an asset in the context of the proposed in the literature possible connection between the mass discrepancy and amount of the near-core mixing in the form of the convective core overshoot \citep[e.g.,][]{Guinan2000,Massey2012,Morrell2014,Claret_Torres2016,Claret_Torres2017,Claret_Torres2018,Claret_Torres2019}. The fact that the present sample contains binaries with and without pulsating components will allow us to study the role of pulsations in the mass discrepancy problem, methodologically (i.e. to what extent the presence of intrinsic variability alters inference of the absolute stellar dimensions from a light curve) and physics wise (e.g. to what extent a wave-induced mixing alters interior and atmospheric properties of stars in models that are used to quantify the mass discrepancy). Moreover, detection of stellar oscillations in at least one of the binary components will enable an independent measurement of the amount of the near-core mixing with asteroseismic methods. A further extension of the sample towards inclusion of more evolved stars will allow us to investigate a possible connection between the mass discrepancy and evolutionary stage of the star, in particular the role of turbulent and radiative pressure terms in the inference of the effective temperature of the star when its surface gravity is accurately known from a photodynamical model of the system \citep[e.g.,][]{Tkachenko2020}.

Summarising, the 15\% sub-sample of the total of 573 eclipsing binary systems that we are currently monitoring spectroscopically, represents one of the largest (if not the largest) eclipsing binary star ensembles that are being observed in a largely consistent way (i.e. with the same ground-based instruments for spectroscopic data and TESS space-based mission for photometric data sets) and that will be analysed by us with a well-established modelling framework described in detail in \citet{Tkachenko2020}. The high level of consistency in the observational strategy and modelling approach is in particular important to minimise otherwise numerous and hardly traceable systematic uncertainties in the investigation of the mass discrepancy problem. Coupled with a homogeneous and in the near future complete coverage of the parameter space, we can conclude that our sample of eclipsing binaries has all the potential to become the foundation to quantify and reesolve the mass discrepancy problem in eclipsing binaries.

Finally, we stress an appreciable capability of the spectral disentangling method to infer precise and accurate spectroscopic orbital elements and individual components' spectra from as little as 6-8 orbital phase-resolved spectroscopic observations. Indeed, using an estimate of the orbital period of the system as the only input to the planning of our observational campaign, we demonstrate that a stable SPD solution can be obtained for the majority of binaries in the sample. This is opposed to the established thinking in the community that an extensive time-series of typically a couple of dozen spectroscopic observations is required for the SPD method to deliver meaningful results. In particular, by comparing our orbital solutions reported in Table~\ref{Tab:SPDOrbit} with those obtained by other research groups based on typically more extended datasets, we find a good agreement for all but a few systems (cf. Section~\ref{sec:discussion}) in the sample. Moreover, systems like AE Pic or AW Cam that are known as SB1 in the literature are discovered by us to be double-lined binaries and thus become ideal candidates for a detailed modelling with the goal to infer absolute dimensions of both binary components. Furthermore, the precision with which we infer the RV semi-amplitudes of both binary components is often comparable to that reported in dedicated studies in the literature \citep[e.g.,][for the AH~Cep and $\alpha$~CrB systems, respectively]{Pavlovski2018,Schmitt2016}. This in turn translates into our ability to infer absolute dimensions of stars with the precision and accuracy better than 3\% when a complementary high-quality ground- or space-based photometric dataset is available (which is the case for all systems in our sample). 

The above-mentioned accuracy of 3\% in stellar mass is sufficiently high to provide a pertinent calibration of SSE models for scientific exploitation of future space-based missions like PLATO \citep{Rauer2014}. Indeed, \citet{Chaplin2014} and \citet{SilvaAguirre2017} demonstrate asteroseismically for solar-like pulsators (i.e. main-sequence Sun-like stars and evolved intermediate-mass stars on the red giant brunch) that $\sim$10\% precision in age can be achieved when the mass and radius of the star are measured with the precision of 4\% and 2\%, respectively. The 10\% age accuracy is one of the most fundamental science requirements of the PLATO mission to be able to characterise an Earth-like planet in the habitable zone of a Sun-like star \citep{Rauer2014}. In this work, we demonstrate that our approach to the planning of spectroscopic observations and subsequent analysis of the obtained data precision and accuracy wise is compliant even with the most stringent requirements of space-based missions such as PLATO.

In the forthcoming papers, we will present detailed analyses of individual systems based on their combined spectroscopic and TESS photometric data. All systems will be studied in the context of the mass discrepancy problem presented in detail in Section~\ref{sec:introduction}. Also, the forthcoming papers will include updates on our ongoing spectroscopic monitoring of the entire sample of 573 candidate eclipsing binaries, in a format similar to that used in the present study.

\begin{acknowledgements}

The research leading to these results has received funding from the KU Leuven Research Council (grant C16/18/005: PARADISE), from the Research Foundation Flanders (FWO) under grant agreements G089422N (NS, AT), V506223N (KP, Scientific stay in Flanders), 11E5620N (SG, PhD Aspirant mandate), 1124321N (LIJ, PhD Aspirant mandate), and 1286521N (DMB, Senior Postdoctoral Fellowship), as well as from the BELgian federal Science Policy Office (BELSPO) through PRODEX grant PLATO. DMB gratefully acknowledges the Engineering and Physical Sciences Research Council (EPSRC) of UK Research and Innovation (UKRI) in the form of a Frontier Research grant under the UK government’s ERC Horizon Europe funding guarantee (grant number [EP/Y031059/1]), and a University Research Fellowship from the Royal Society (grant number: [URF{\textbackslash}R1{\textbackslash}231631]). 

Based on observations made with the Mercator Telescope, operated on the island of La Palma by the Flemish Community, at the Spanish Observatorio del Roque de los Muchachos of the Instituto de Astrofísica de Canarias. Based on observations obtained with the HERMES spectrograph, which is supported by the Research Foundation - Flanders (FWO), Belgium, the Research Council of KU Leuven, Belgium, the Fonds National de la Recherche Scientifique (F.R.S.-FNRS), Belgium, the Royal Observatory of Belgium, the Observatoire de Genève, Switzerland and the Thüringer Landessternwarte Tautenburg, Germany.

Some of the observations used in this work were obtained with the FEROS spectrograph mounted on the 2.2-m MPG/ESO telescope at the ESO La Silla observatory under program 0106.A-0906(A).

The TESS data in this paper were obtained from the Mikulski Archive for Space Telescopes (MAST) at the Space Telescope Science Institute (STScI), which is operated by the Association of Universities for Research in Astronomy, Inc., under NASA contract NAS5-26555. Support to MAST for these data is provided by the NASA Office of Space Science via grant NAG5-7584 and by other grants and contracts. Funding for the TESS mission is provided by the NASA Explorer Program. 

This research has made use of the SIMBAD database, operated at CDS, Strasbourg, France; the SAO/NASA Astrophysics Data System; and the VizieR catalog access tool, CDS, Strasbourg, France.

\end{acknowledgements}

\bibliographystyle{aa}
\typeout{}
\bibliography{mybib.bib}

\begin{appendix}
\section{Classification of the sample stars}
In this section, we present our spectroscopic and photometric classifications for all stars included in the present sample. 

\onecolumn
\addtolength{\tabcolsep}{-3pt}
\begin{ThreePartTable}
\begin{TableNotes}
\item [] ``N'' refers to the number of spectra we have at our disposal. Period uncertainty is provided in parentheses in terms of the last digit.
\item [1] Eclipse signal is detected in the TESS data but contamination as the source cannot be excluded.
\item [2] Variable eclipse depth such that eclipses practically disappear in the TESS Sectors 61 \& 62.
\end{TableNotes}
\begin{longtable}{lllllll}
    \caption{Spectroscopic and photometric classification for stars studied in this work.}
    \label{Tab:targets} \\
    \hline\hline
   \multicolumn{3}{c}{Star ID} & \multicolumn{1}{c}{Period} & \multirow{2}{*}{N} & \multicolumn{2}{l}{Classification}\\
    Main & Alternative & TIC & \multicolumn{1}{c}{(d)} & & Spec. & Phot.\\
    \hline
    \endfirsthead
    \caption{continued.} \\
    \hline\hline
   \multicolumn{3}{c}{Star ID} & \multicolumn{1}{c}{Period} & \multirow{2}{*}{N} & \multicolumn{2}{l}{Classification}\\
    Main & Alternative & TIC & \multicolumn{1}{c}{(d)} & & Spec. & Phot. \\
    \hline
    \endhead
    \hline
    \insertTableNotes
    \endfoot
\multicolumn{7}{c}{{\sc hermes} sample} \\
BD+13 1880 &   BD+13 1880 & TIC 186812530 & 3.677(2)  & 10 & SB1 & EB\\
BD+30 3184   & BD+30 3184 & TIC 23470753  & 4.237216(5)  & 8  & SB2 & EB, SPB or $\delta$~Sct\\   
BD+36 3317   & BD+36 3317 & TIC 237195907 & 4.302147(3)  & 8  & SB2 & EB, $\delta$~Sct\\ 
BD+47 1906   & BD+47 1906 & TIC 253049152 & 1.3993(4)  & 12 & SB2 & EB\\   
BD+67 1049   & BD+67 1049 & TIC 219110814 & 3.5888926(6)  & 8  & SB1 & EB\\   
65 UMa B\tnote{1}   & HD 103498 & 	TIC 141148944 & 0.8652(2)  & 11 & single & $\delta$~Sct\\   
HD 107379    & HD 107379 & 	TIC 148895442 & 3.595266(3)  & 9  & SB2 & EB\\   
EL CVn    & HD 116608 & 	TIC 165371937 & 0.795624(8)  & 11 & SB2 & EB\\   
NO Dra    & HD 135437 & 	TIC 202442974 & 2.738920(2)  & 9 & SB2 & EB\\
HD 135466\tnote{1}    & HD 135466 & 	TIC 202442982 & 29.26(2)  & 8  & LPV & \\  
HD 138305    & HD 138305 & 	TIC 368291074 & 3.498(3)  & 8 & SB2  & EB, RotMod or ellipsoidal\\  
$\alpha$ CrB    & HD 139006 & 	TIC 274945059 & 17.360(5) & 14 & SB2 & EB\\    
HD 13970     & HD 13970	  & TIC 264614791 & 3.508808(2)  & 9  & SB2 & EB, $\beta$~Cep\\  
HD 150781A   & HD 150781a & TIC 349444267 & 7.457(5)  & 8  & SB2 & EB, $\gamma$~Dor\\   
V920 Her  & HD 151972 & 	TIC 143009538 & 6.926(3)  & 8 & SB2  & EB\\  
HD 158148    & HD 158148 & 	TIC 351701483 & 17.91032(2) & 9  & SB2 & EB, SPB\\    
HD 160363    & HD 160363 & 	TIC 311433319 & 1.837710(2)  & 11 & SB1 & EB\\   
V994 Her  & HD 170314 & 	TIC 424508303 & 2.083296(1)  & 9  & SB4 & EB, high-order multiple \\  
HD 172133    & HD 172133 & 	TIC 8705972   & 3.598199(4)	  & 8  & SB1 & EB, RotMod or ellipsoidal\\  
BH Dra    & HD 178001 & 	TIC 377192659 & 1.8172380(2)  & 9  & SB2 & EB\\   
V2108 Cyg & HD 191530 & 	TIC 378395625 & 2.560339(1)  & 7  & SB2 & EB\\   
V1898 Cyg & HD 200776 & 	TIC 273173532 & 1.5131193(2)  & 42 & SB2 & EB\\   
GK Cep    & HD 205372 & 	TIC 256352113 & 0.9361715(2)  & 7  & SB3 & EB\\   
V383 Cep  & HD 208106 & 	TIC 410522328 & 1.4957437(3)  & 11 & SB2 & EB\\   
HD 208510    & HD 208510 & 	TIC 299494754 & 1.5970321(2)  & 8  & SB2 & EB, RotMod or ellipsoidal\\   
AH Cep    & HD 216014 & 	TIC 377506471 & 1.7747483(4)  & 11 & SB2 & EB\\   
V350 And  & HD 2189	 & 	TIC 58107375  & 1.7111454(2) & 8  & SB2	 & EB, $\gamma$~Dor/$\delta$~Sct hybrid\\    
HD 234650    & HD 234650 & 	TIC 21189379  & 7.590363(6) & 8  & SB3 & EB\\ 	   
u Tau      & HD 23466	  & TIC 426588729 & 2.4245(9) & 14 & SB2 & EB, RotMod or SPB\\   
HD 234713    & HD 234713 & 	TIC 48087401  & 3.067219(3) & 8  & SB2 & EB\\ 	   
HD 237866    & HD 237866 & 	TIC 137905382 & 1.5238(4)  & 9  & SB1 & EB\\  
HD 246047\tnote{1}    & HD 246047 & 	TIC 116331699 & 9.91493(2) & 11 & SB1 or LPV & \\ 	   
IM Aur    & HD 33853	  & TIC 368180294 & 1.2472686(2) & 11 & SB2 & EB\\ 	  
HD 34382\tnote{1}     & HD 34382	  & TIC 2234723   & 2.4619009(4) & 13 & LPV & RotMod\\ 	  
HD 348568    & HD 348568 & 	TIC 342480862 & 11.727766(4) & 8  & SB2 & EB\\  	   
HD 348725    & HD 348725 & 	TIC 342794723 & 1.531346(1) & 7  & SB2 & EB\\ 	   
HD 350685    & HD 350685 & 	TIC 392053854 & 2.873796(2) & 8  & SB1 & EB, $\beta$~Cep\\ 	   
HD 37646     & HD 37646	  & TIC 75507062  & 0.6726158(4) & 8  & SB1 & EB\\ 	  
WW Aur    & HD 46052	  & TIC 172171873 & 2.525018(1) & 9 & SB2 & EB\\
V459 Aur  & HD 46552	  & TIC 172421004 & 1.0626466(7)  & 14 & SB2 & EB\\  
HD 47046     & HD 47046	  & TIC 353759550 & 1.7166(6)  & 9  & SB2 & EB, RotMod and ellipsoidal\\  
AW Cam    & HD 48049	  & TIC 456263109 & 0.77134620(7)  & 8  & SB2 & EB\\  
HD 54159     & HD 54159	  & TIC 367567347 & 0.80858(3)  & 9  & single & RotMod\\  
HD 57158     & HD 57158	  & TIC 302907601 & 1.6982479(7)  & 8  & SB4 & EB, high-order multiple?\\ 
HD 63887     & HD 63887	  & TIC 457101125 & 3.791285(3)  & 8  & SB2 & EB\\ 
V766 Cas  & HD 8027	 & 	TIC 241017747 & 2.3296557(4)  & 9  & SB2 & EB\\  
HD 89601     & HD 89601	  & TIC 150251466 & 12.388(7) & 9 & SB1 & EB\\   
TYC 3529-2494-1\tnote{1} & TYC 3529-2494-1 & TIC 7694914 & 0.4450381(2) & 9 & LPV & RotMod\\
V453 Cyg  & HD 227696 & TIC 90349611  & 3.890(1)  & 28 & SB2 & EB, $\beta$~Cep\\
NY Cep	 &  HD 217312 & TIC 13389059  & 15.2884(2) & 8 & SB2 & EB \\
16 Lac	 &  HD 216916 & TIC 129538133 & 12.097(1) & 180 & SB1 & EB, $\beta$~Cep\\
V436 Per &  HD 11241 & TIC 403625251 & 25.936(1) & 41 & SB2 & EB, SLF\\
RX Dra	 &  RX Dra  & TIC 377190161 & 3.876(1) & 26 & SB2 & EB, $\gamma$~Dor\\
V1425 Cyg & HD 202000 & TIC 344456744 & 1.2523886(1) & 44 & SB2 & EB\\
V446 Cep &  HD 210478 & TIC 335265326 & 3.808(1) & 73 & SB2 & EB, $\beta$~Cep\\
CM Lac	 &  HD 209147 & TIC 331221558 & 1.6047(1) & 27 & SB2 & EB, $\gamma$~Dor\\
V398 Lac &  HD 210180 & TIC 326374705 & 5.406081(1) & 26 & SB2 & EB\\
V402 Lac &  HD 210405 & TIC 468792279 & 3.7818043(8) & 30 & SB2 & EB, SPB\\ 
\multicolumn{7}{c}{{\sc feros} sample} \\
HD 100737\tnote{1} & HD100737 & TIC 290391601 &  2.552548(2) & 9 & SB1 & $\beta$~Cep?\\
HD 104233 & HD104233 & TIC 307687961 &        1.823975(1) & 10 & SB2 & EB\\
HD 121776 & HD121776 & TIC 448375181 & 1.737896(1) & 9 & SB2 & EB\\
HD 28913  & HD28913 & TIC 170729895 &         1.4906780(1) & 7 & LPV & RotMod\\
HD 300344 & HD300344 & TIC 274687574 &        2.7902916(3) & 7 & SB2 & EB, RotMod\\
HD 304241 & HD304241 & TIC 451259413 &        2.730422(3) & 9 & SB1 & EB\\
HD 309317 & HD309317 & TIC 306139502 &        2.257301(2) & 9 & SB2 & EB, $\beta$~Cep\\
AE Pic & HD46792 & TIC 150442264 &         2.9816923(9) & 8 & SB2 & EB, RotMod and $\beta$~Cep\\
HD 51981\tnote{1}  & HD51981 & TIC 147314529 & 0.4627711(2) & 7 & single & \\
HD 52349  & HD52349 & TIC 80041531  &         2.775849(3) & 8 & SB2 & EB, SPB \\
V386 Pup & HD62738 & TIC 175254818 & 1.649314(1) & 6 & SB2 & EB \\
HD 66235 &  HD66235 & TIC 285413162 &         1.5941375(1) & 7 & LPV & RotMod\\
HD 67025 &  HD67025 & TIC 79935432  &         1.2823665(1) & 7 & SB2 & EB \\
HD 68340\tnote{1} &  HD68340 & TIC 145405941 &         1.3069152(1) & 7 & single & \\
HD 75872 &  HD75872 & TIC 29216374  &         0.94508671(6) & 6 & SB1 & EB \\
HD 79365\tnote{1} &  HD79365 & TIC 74715631  &         0.90672432(6) & 8 & LPV & \\
HD 82110 &  HD82110 & TIC 438089724 &         1.8810195(2) & 8 & SB2 & EB, SPB \\
HD 84493 &  HD84493 & TIC 363146191 &         6.8760(2) & 8 & SB1 & EB, SLF \\
HD 91141 &  HD91141 & TIC 457540424 &         2.3821886(4) & 10 & SB2 & EB\\
HD 91154 &  HD91154 & TIC 457545293 &         3.6631544(5) & 10 & SB1 & EB \\
HD 92741\tnote{1} &  HD92741 & TIC 458561474 &         5.372886(4) & 10 & LPV & $\beta$~Cep? \\
HD 97966\tnote{1} &  HD97966 & TIC 450276745 &         1.2666420(8) & 9 & single & \\
CD-45 4393\tnote{2} & TYC8151-937-1 & TIC 141858108 &  2.0494343(2) & 6 & SB1 & EB \\
CD-47 4364\tnote{1} & TYC8155-1212-1 & TIC 270844716 & 3.6580419(5) & 8 & SB1 or LPV & SPB \\
CD-56 1160 & TYC8514-106-1 & TIC 382044531 &  0.7903919(1) & 6 & SB2 & EB \\
\end{longtable}
\end{ThreePartTable}

\section{SPD-based orbital solutions}
In this section we present SPD-based orbital solutions for all stars that are classified as either SB1 or SB2 systems in Table~\ref{Tab:targets}.

\begin{ThreePartTable}
\begin{TableNotes}
\item [] $e$, $\omega$, and $K_{\rm i}$ stand for the eccentricity, time of periastron passage, and RV semi-amplitudes, respectively. Cases where eccentricity and argument of periastron were fixed to 0.0 and 90$^\circ$, respectively, are indicated with the superscript ``f''. Orbital period values and their uncertainties are adopted from Table~\ref{Tab:targets}.
\item [1] The star is classified as an SB2 system based on the visual inspection of the LSD profiles and/or original observed spectra, however the SPD solution could be obtained for the brighter primary component only.
\item [2] Same as above but no stable SPD solution could be obtained for either of the components.
\item [3] Either an SB1 system with a low $K$ semi-amplitude of about 1.5~\kms\ or a single star showing LPVs: none of the methods employed in this work favours any of those hypothesis.
\end{TableNotes}
\begin{longtable}{llllllllllll}
    \caption{Spectroscopic orbital elements for all SB1 and SB2 systems studied in this work.}
    \label{Tab:SPDOrbit} \\
    \hline\hline
   \multicolumn{3}{c}{Star ID} & Period & \multicolumn{2}{c}{$e$} & \multicolumn{2}{c}{$\omega$} & \multicolumn{2}{c}{$K_{\rm 1}$} & \multicolumn{2}{c}{$K_{\rm 2}$} \\
   & & & & value & error & value & error & value & error & value & error\\
    Main & Alt. & TIC & \multicolumn{1}{c}{(d)} & & & \multicolumn{2}{c}{(degrees)} & \multicolumn{4}{c}{(\kms)} \\
    \hline
    \endfirsthead
    \caption{continued.} \\
    \hline\hline
   \multicolumn{3}{c}{Star ID} & Period & \multicolumn{2}{c}{$e$} & \multicolumn{2}{c}{$\omega$} & \multicolumn{2}{c}{$K_{\rm 1}$} & \multicolumn{2}{c}{$K_{\rm 2}$} \\
   & & & & value & error & value & error & value & error & value & error\\
    Main & Alt. & TIC & \multicolumn{1}{c}{(d)} & & & \multicolumn{2}{c}{(degrees)} & \multicolumn{4}{c}{(\kms)} \\
    \hline
    \endhead
    \hline
    \insertTableNotes
    \endfoot
\multicolumn{12}{c}{{\sc hermes} sample} \\
BD+13 1880 &   BD+13 1880 & TIC 186812530 & 3.677(2)  & 0.0$^{\rm f}$ & 0.0$^{\rm f}$ & 90$^{\rm f}$ & 0.0$^{\rm f}$ & 5.0 & 0.1 & --- & --- \\
BD+30 3184 &   BD+30 3184 & TIC 23470753 & 4.237216(5)  & 0.20 & 0.01 & 121.9 & 5.1 & 82.7 & 1.2 & 116.8 & 3.2 \\
BD+36 3317 &   BD+36 3317 & TIC 237195907 & 4.302147(3)  & 0.0$^{\rm f}$ & 0.0$^{\rm f}$ & 90$^{\rm f}$ & 0.0$^{\rm f}$ & 82.93 & 0.03 & 127.3 & 0.1 \\
BD+47 1906 &   BD+47 1906 & TIC 253049152 & 1.3993(4)  & 0.0$^{\rm f}$ & 0.0$^{\rm f}$ & 90$^{\rm f}$ & 0.0$^{\rm f}$ & 20.1 & 0.5 & 213.6 & 3.2\\
BD+67 1049 &   BD+67 1049 & TIC 219110814 & 3.5888926(6)  & 0.0$^{\rm f}$ & 0.0$^{\rm f}$ & 90$^{\rm f}$ & 0.0$^{\rm f}$ & 18.64 & 0.06 & --- & ---\\
HD 107379 &   HD 107379 & TIC 148895442 & 3.595266(3)  & 0.0$^{\rm f}$ & 0.0$^{\rm f}$ & 90$^{\rm f}$ & 0.0$^{\rm f}$ & 14.9 & 0.8 & 76.4 & 0.6\\
EL CVn &   HD 116608 & TIC 165371937 & 0.795624(8)  & 0.0$^{\rm f}$ & 0.0$^{\rm f}$ & 90$^{\rm f}$ & 0.0$^{\rm f}$ & 30.0 & 1.6 & 242.4 & 2.7\\
NO Dra &   HD 135437 & TIC 202442974 & 2.738920(2)  & 0.02 & 0.04 & 217 & 11 & 77.8 & 0.5 & 154.9 & 3.0\\
HD 138305 &   HD 138305 & TIC 368291074 & 3.498(3)  & 0.0$^{\rm f}$ & 0.0$^{\rm f}$ & 90$^{\rm f}$ & 0.0$^{\rm f}$ & 106.4 & 0.2 & 109.39 & 0.08\\
$\alpha$ CrB &   HD 139006 & TIC 274945059 & 17.360(5)  & 0.3795 & 0.0002 & 314.7 & 0.8 & 37.1 & 1.3 & 98.5 & 0.8\\
HD 13970 &   HD 13970 & TIC 264614791 & 3.508808(2)  & 0.0$^{\rm f}$ & 0.0$^{\rm f}$ & 90$^{\rm f}$ & 0.0$^{\rm f}$ & 117.1 & 1.4 & 188.2 & 4.3\\
HD 150781A &   HD 150781A & TIC 349444267 & 7.457(5)  & 0.139 & 0.001 & 189.0 & 0.6 & 84.8 & 0.6 & 85.7 & 0.1\\
V920 Her &   HD 151972 & TIC 143009538 & 6.926(3)  & 0.0$^{\rm f}$ & 0.0$^{\rm f}$ & 90$^{\rm f}$ & 0.0$^{\rm f}$ & 85.4 & 0.4 & 90.98 & 0.09\\
HD 158148    & HD 158148 & 	TIC 351701483 & 17.91032(2)  & 0.0$^{\rm f}$ & 0.0$^{\rm f}$ & 90$^{\rm f}$ & 0.0$^{\rm f}$ & 18.0 & 1.1 & 129.0 & 0.4\\
HD 160363    & HD 160363 & 	TIC 311433319 & 1.837710(2)  & 0.0$^{\rm f}$ & 0.0$^{\rm f}$ & 90$^{\rm f}$ & 0.0$^{\rm f}$ & 28.2 & 0.8 & --- & ---\\
HD 172133    & HD 172133 & 	TIC 8705972 & 3.598199(4)  & 0.0$^{\rm f}$ & 0.0$^{\rm f}$ & 90$^{\rm f}$ & 0.0$^{\rm f}$ & 43.97 & 0.04 & --- & ---\\
BH Dra    & HD 178001 & 	TIC 377192659 & 1.8172380(2)  & 0.0$^{\rm f}$ & 0.0$^{\rm f}$ & 90$^{\rm f}$ & 0.0$^{\rm f}$ & 105.8 & 0.6 & 162.8 & 2.0\\
V2108 Cyg & HD 191530 & 	TIC 378395625 & 2.560339(1)  & 0.0$^{\rm f}$ & 0.0$^{\rm f}$ & 90$^{\rm f}$ & 0.0$^{\rm f}$ & 44.7 & 3.2 & 72.6 & 1.4\\
V1898 Cyg & HD 200776 & 	TIC 273173532 & 1.5131193(2)  & 0.0$^{\rm f}$ & 0.0$^{\rm f}$ & 90$^{\rm f}$ & 0.0$^{\rm f}$ & 54.6 & 1.5 & 293.0 & 1.4\\
V383 Cep  & HD 208106 & 	TIC 410522328 & 1.4957437(3) & 0.0$^{\rm f}$ & 0.0$^{\rm f}$ & 90$^{\rm f}$ & 0.0$^{\rm f}$ & 122.9 & 0.7 & 151.8 & 4.2\\
HD 208510    & HD 208510 & 	TIC 299494754 & 1.5970321(2) & 0.0$^{\rm f}$ & 0.0$^{\rm f}$ & 90$^{\rm f}$ & 0.0$^{\rm f}$ & 68.0 & 6.1 & 243.7 & 5.6\\
AH Cep    & HD 216014 & 	TIC 377506471 & 1.7747483(4) & 0.0$^{\rm f}$ & 0.0$^{\rm f}$ & 90$^{\rm f}$ & 0.0$^{\rm f}$ & 230.1 & 2.0 & 277.1 & 3.9\\
V350 And  & HD 2189	 & 	TIC 58107375 & 1.7111454(2) & 0.0$^{\rm f}$ & 0.0$^{\rm f}$ & 90$^{\rm f}$ & 0.0$^{\rm f}$ & 125.5 & 1.0 & 133.7 & 1.3\\
u Tau      & HD 23466	  & TIC 426588729 & 2.4245(9) & 0.22 & 0.01 & 230 & 8 & 22.5 & 2.7 & 75.2 & 2.8\\
HD 234713    & HD 234713 & 	TIC 48087401 & 3.067219(3) & 0.0$^{\rm f}$ & 0.0$^{\rm f}$ & 90$^{\rm f}$ & 0.0$^{\rm f}$ & 101.87 & 0.03 & 102.9 & 0.6\\
HD 237866    & HD 237866 & 	TIC 137905382 & 1.5238(4) & 0.0$^{\rm f}$ & 0.0$^{\rm f}$ & 90$^{\rm f}$ & 0.0$^{\rm f}$ & 19.5 & 1.8 & --- & ---\\
HD 246047\tnote{3}    & HD 246047 & 	TIC 116331699 & 9.91493(2) & 0.0$^{\rm f}$ & 0.0$^{\rm f}$ & 90$^{\rm f}$ & 0.0$^{\rm f}$ & 1.3 & 0.6 & --- & ---\\
IM Aur    & HD 33853	  & TIC 368180294 & 1.2472686(2) & 0.0$^{\rm f}$ & 0.0$^{\rm f}$ & 90$^{\rm f}$ & 0.0$^{\rm f}$ & 67.8 & 0.6 & 285.8 & 1.0\\
HD 348568    & HD 348568 & 	TIC 342480862 & 11.727766(4) & 0.0$^{\rm f}$ & 0.0$^{\rm f}$ & 90$^{\rm f}$ & 0.0$^{\rm f}$ & 23.8 & 2.0 & 131.3 & 3.8\\
HD 348725    & HD 348725 & 	TIC 342794723 & 1.531346(1) & 0.0$^{\rm f}$ & 0.0$^{\rm f}$ & 90$^{\rm f}$ & 0.0$^{\rm f}$ & 115.4 & 2.5 & 143.9 & 3.1\\
HD 350685    & HD 350685 & 	TIC 392053854 & 2.873796(2) & 0.0$^{\rm f}$ & 0.0$^{\rm f}$ & 90$^{\rm f}$ & 0.0$^{\rm f}$ & 34.1 & 2.9 & --- & ---\\
HD 37646     & HD 37646	  & TIC 75507062 & 0.6726158(4) & 0.0$^{\rm f}$ & 0.0$^{\rm f}$ & 90$^{\rm f}$ & 0.0$^{\rm f}$ & 28.7 & 0.1 & --- & ---\\
WW Aur    & HD 46052	  & TIC 172171873 & 2.525018(1) & 0.0$^{\rm f}$ & 0.0$^{\rm f}$ & 90$^{\rm f}$ & 0.0$^{\rm f}$ & 117.53 & 0.02 & 124.59 & 0.04\\
V459 Aur  & HD 46552	  & TIC 172421004 & 1.0626466(7) & 0.0$^{\rm f}$ & 0.0$^{\rm f}$ & 90$^{\rm f}$ & 0.0$^{\rm f}$ & 65.2 & 1.6 & 149.8 & 2.5\\
HD 47046     & HD 47046	  & TIC 353759550 & 1.7166(6) & 0.0$^{\rm f}$ & 0.0$^{\rm f}$ & 90$^{\rm f}$ & 0.0$^{\rm f}$ & 14.0 & 0.7 & 134.5 & 1.3\\
AW Cam    & HD 48049	  & TIC 456263109 & 0.77134620(7) & 0.0$^{\rm f}$ & 0.0$^{\rm f}$ & 90$^{\rm f}$ & 0.0$^{\rm f}$ & 113.5 & 2.7 & 252.9 & 3.3\\
HD 63887     & HD 63887	  & TIC 457101125 & 3.791285(3) & 0.0$^{\rm f}$ & 0.0$^{\rm f}$ & 90$^{\rm f}$ & 0.0$^{\rm f}$ & 104.1 & 0.4 & 119.2 & 0.9\\
V766 Cas  & HD 8027	 & 	TIC 241017747 & 2.3296557(4) & 0.09 & 0.02 & 232 & 17 & 129.9 & 1.0 & 154.4 & 8.5\\
HD 89601     & HD 89601	  & TIC 150251466 & 12.388(7) & 0.4207 & 0.0003 & 191.28 & 0.08 & 30.60 & 0.01 & --- & ---\\
V453 Cyg  & HD 227696 & TIC 90349611 & 3.890(1) & 0.027 & 0.002 & 151.3 & 11.3 & 173.7 & 1.0 & 221.4 & 1.0\\
NY Cep	 &  HD 217312 & TIC 13389059 & 15.2884(2) & 0.444 & 0.001 & 54.2 & 1.7 & 110.6 & 1.6 & 141.7 & 7.0\\
16 Lac	 &  HD 216916 & TIC 129538133 & 12.097(1) & 0.047 & 0.004 & 40.5 & 1.6 & 23.75 & 0.06 & --- & ---\\
V436 Per &  HD 11241 & TIC 403625251 & 25.936(1) & 0.373 & 0.009 & 109.5 & 1.0 & 94.1 & 1.8 & 99.9 & 1.0\\
RX Dra	 &  RX Dra  & TIC 377190161 & 3.876(1) & 0.0$^{\rm f}$ & 0.0$^{\rm f}$ & 90$^{\rm f}$ & 0.0$^{\rm f}$ & 85.8 & 0.2 & 104.6 & 0.3\\
V1425 Cyg & HD 202000 & TIC 344456744 & 1.2523886(1) & 0.0$^{\rm f}$ & 0.0$^{\rm f}$ & 90$^{\rm f}$ & 0.0$^{\rm f}$ & 143.8 & 1.3 & 218.4 & 2.2\\
V446 Cep &  HD 210478 & TIC 335265326 & 3.808(1) & 0.0150 & 0.0005 & 69.7 & 17.4 & 42.3 & 1.9 & 309.7 & 2.4\\
CM Lac	 &  HD 209147 & TIC 331221558 & 1.6047(1) & 0.0$^{\rm f}$ & 0.0$^{\rm f}$ & 90$^{\rm f}$ & 0.0$^{\rm f}$ & 121.0 & 0.2 & 155.3 & 0.8\\
V398 Lac &  HD 210180 & TIC 326374705 & 5.406081(1) & 0.23 & 0.01 & 233.6 & 7.5 & 90.7 & 1.5 & 139.3 & 2.6\\
V402 Lac &  HD 210405 & TIC 468792279 & 3.7818043(8) & 0.358 & 0.005 & 56.7 & 2.9 & 125.5 & 0.3 & 128.5 & 0.5\\
\multicolumn{12}{c}{{\sc feros} sample} \\
HD 100737 & HD100737 & TIC 290391601 & 2.552548(2) & 0.0$^{\rm f}$ & 0.0$^{\rm f}$ & 90$^{\rm f}$ & 0.0$^{\rm f}$ & 56.9 & 0.5 & --- & ---\\
HD 104233 & HD104233 & TIC 307687961 & 1.823975(1) & 0.0$^{\rm f}$ & 0.0$^{\rm f}$ & 90$^{\rm f}$ & 0.0$^{\rm f}$ & 42.0 & 0.9 & 92.3 & 4.2\\
HD 121776\tnote{1} & HD121776 & TIC 448375181 & 1.737896(1) & 0.0$^{\rm f}$ & 0.0$^{\rm f}$ & 90$^{\rm f}$ & 0.0$^{\rm f}$ & 81.8 & 1.1 & --- & ---\\
HD 300344\tnote{2} & HD300344 & TIC 274687574 & 2.7902916(3) & --- & --- & --- & --- & --- & --- & --- & ---\\
HD 304241 & HD304241 & TIC 451259413 & 2.730422(3) & 0.0$^{\rm f}$ & 0.0$^{\rm f}$ & 90$^{\rm f}$ & 0.0$^{\rm f}$ & 84.6 & 1.3 & --- & ---\\
HD 309317 & HD309317 & TIC 306139502 & 2.257301(2) & 0.0$^{\rm f}$ & 0.0$^{\rm f}$ & 90$^{\rm f}$ & 0.0$^{\rm f}$ & 34.5 & 0.3 & 78.7 & 1.4\\
AE Pic & HD46792 & TIC 150442264 & 2.9816923(9) & 0.075 & 0.005 & 58.4 & 26.4 & 114.1 & 1.3 & 158.9 & 3.6\\
HD 52349\tnote{1}  & HD52349 & TIC 80041531 & 2.775849(3) & 0.37 & 0.01 & 312.6 & 7.3 & 54.8 & 0.6 & --- & ---\\
V386 Pup\tnote{2} & HD62738 & TIC 175254818 & 1.649314(1) & --- & --- & --- & --- & --- & --- & --- & ---\\
HD 67025 &  HD67025 & TIC 79935432 & 1.2823665(1) & 0.0$^{\rm f}$ & 0.0$^{\rm f}$ & 90$^{\rm f}$ & 0.0$^{\rm f}$ & 147.4 & 1.9 & 173.7 & 2.7\\
HD 75872 &  HD75872 & TIC 29216374 & 0.94508671(6) & 0.0$^{\rm f}$ & 0.0$^{\rm f}$ & 90$^{\rm f}$ & 0.0$^{\rm f}$ & 45.9 & 1.0 & --- & ---\\
HD 82110 &  HD82110 & TIC 438089724 & 1.8810195(2) & 0.0$^{\rm f}$ & 0.0$^{\rm f}$ & 90$^{\rm f}$ & 0.0$^{\rm f}$ & 78.2 & 1.0 & 178.8 & 1.5\\
HD 84493 &  HD84493 & TIC 363146191 & 6.8760(2) & 0.0$^{\rm f}$ & 0.0$^{\rm f}$ & 90$^{\rm f}$ & 0.0$^{\rm f}$ & 29.20 & 0.06 & --- & ---\\
HD 91141 &  HD91141 & TIC 457540424 & 2.3821886(4) & 0.0$^{\rm f}$ & 0.0$^{\rm f}$ & 90$^{\rm f}$ & 0.0$^{\rm f}$ & 15.58 & 0.08 & 72.7 & 1.8\\
HD 91154 &  HD91154 & TIC 457545293 & 3.6631544(5) & 0.0$^{\rm f}$ & 0.0$^{\rm f}$ & 90$^{\rm f}$ & 0.0$^{\rm f}$ & 36.5 & 1.5 & --- & ---\\
CD-45 4393 & TYC8151-937-1 & TIC 141858108 & 2.0494343(2) & 0.0$^{\rm f}$ & 0.0$^{\rm f}$ & 90$^{\rm f}$ & 0.0$^{\rm f}$ & 16.9 & 0.7 & --- & ---\\
CD-47 4364\tnote{3} & TYC8155-1212-1 & TIC 270844716 & 3.6580419(5) & 0.0$^{\rm f}$ & 0.0$^{\rm f}$ & 90$^{\rm f}$ & 0.0$^{\rm f}$ & 1.58 & 0.09 & --- & ---\\
CD-56 1160 & TYC8514-106-1 & TIC 382044531 & 0.7903919(1) & 0.0$^{\rm f}$ & 0.0$^{\rm f}$ & 90$^{\rm f}$ & 0.0$^{\rm f}$ & 65.7 & 1.5 & 155.6 & 2.2\\
\end{longtable}
\end{ThreePartTable}

\end{appendix}

\end{document}